\documentclass[sn-mathphys]{sn-jnl}% Math and Physical Sciences Reference Style
\jyear{2021}%

\theoremstyle{thmstyleone}%
\theoremstyle{thmstyletwo}%

\theoremstyle{thmstylethree}%

\usepackage{graphicx}
\usepackage{bm}
\usepackage{xcolor}

\begin{document}

\newcommand{\be}{\begin{equation}}
\newcommand{\ee}[1]{\label{#1}\end{equation}}
\newcommand{\bem}{\begin{eqnarray}}
\newcommand{\eem}[1]{\label{#1}\end{eqnarray}}
\newcommand{\eq}[1]{Eq.~(\ref{#1})}
\newcommand{\Eq}[1]{Equation~(\ref{#1})}
\newcommand{\ua}{\uparrow}
\newcommand{\da}{\downarrow}
\newcommand{\g}{\dagger}

\newcommand{\rc}[1]{\textcolor{red}{#1}}

\title[Andreev states and  SNS junction]{Andreev reflection, Andreev states, and long ballistic SNS junction}


\author*{\fnm{Edouard} \sur{Sonin}}\email{sonin@cc.huji.ac.il}

\affil{\orgdiv{Racah Institute  of Physics}, \orgname{Hebrew University of Jerusalem}, \orgaddress{\street{Givat Ram}, \city{Jerusalem}, \postcode{9190401}, %\state{State},
 \country{Israel}}}


\abstract{The analysis in the present paper is based on the most known concept  introduced by the brilliant physicist  Alexander Andreev: Andreev bound states in a normal metal sandwiched between two  superconductors.  The paper  presents results of  direct calculations of {\em ab initio} expressions for the currents in a long ballistic SNS junction. The expressions are expanded in $1/L$  ($L$ is  the thickness of the normal layer). The main contribution $\propto  1/L$ to  the current agrees with the results obtained in the past, but the analysis suggests a new physical picture of the charge transport through the junction free from the problem  with the charge conservation law. The saw-tooth current-phase relation at $T=0$ directly follows from the Galilean invariance of the Bogolyubov\,\textendash\,de Gennes equations proved in the paper. The proof is valid for any variation of the energy gap in space if the Andreev reflection is the only scattering process. The respective roles of the contributions of bound and continuum states to the current are clarified. They depend on the junction dimensionality.}




\maketitle

%\section{Introduction}\label{sec1}

%\begin{figure}[h]%
%\centering
%\includegraphics[width=0.9\textwidth]{2Dnucl.pdf}
%\caption{Nucleation of a vortex near. }\label{fig1}
%\end{figure}

\section{Introduction} \label{Intr}

%\subsection{Andreev reflection and  Andreev states}

The work of Andreev demonstrating the possibility of Andreev reflection was published nearly sixty years ago \cite{And64}. Remarkably, the interest to this phenomenon is not decreasing but rather increasing with time. More and more applications of it are being suggested. While in normal reflection a particle (quasiparticle) changes  direction of its momentum, in the Andreev reflection  it changes the direction of its group velocity. For reflection from an interface between a normal metal and a superconductor this means that a particle  in a normal metal is reflected as a hole, or {\em vice versa}.

Andreev considered reflection from a planar interface. But it takes  place also for quasiparticle reflection from a quantum vortex and not only in superconductors. Rotons in superfluid $^4$He have a spectrum similar to that of BCS quasiparticles. Andreev reflection from a velocity field around the vortex contributes to  a force  on the vortex from moving rotons in superfluid $^4$He and BCS quasiparticles in superconductors and in superfluid $^3$He. In fact, the first example of the Andreev reflection appeared in the paper by Lifshitz and Pitaevskii \cite{Lif57} on a force on a vortex from rotons even before Andreev's paper  \cite{And64}, although it is difficult to see this in their less-than-half page communication giving  only final expressions for the force. Later calculations of this force with more details and some corrections  were published \cite{Son75,Gal,Kop76,EBS}. Andreev reflection by vortices plays an important role in  investigations of quantum turbulence  in $^3$He, which is a Fermi superfluid described by the BCS theory \cite{Lancast}.  Recently Skrbek and  Sergeev \cite{ScrbSerg} suggested to use Andreev reflection of rotons by vortices in investigations of quantum turbulence  in superfluid $^4$He.

The concept of Andreev reflection naturally brought Andreev to the next fundamental concept of condensed matter (and maybe not  only condensed matter) physics: Andreev bound states \cite{And65}. In a normal metal between two superconductors quantum states with  energies less than the gap in superconductors are not able to penetrate into the superconductor. Quasiparticles at these states jump forth and back with multiple  Andreev reflections from interfaces. After any Andreev reflection a particle becomes a hole, or {\em vice versa}.

Andreev  reflection and Andreev states are key concepts for the problem addressed in the present paper: long ballistic SNS junction (normal metal sandwiched between two  superconductors). It was noticed long ago  \cite{Kulik,Ishii,Bard} that if the normal metal layer is ballistic the Josephson effect exists even for layer thickness much exceeding the coherence length. These works used the self-consistent field method \cite{deGen}.  In this method  an effective pairing potential is introduced, which transforms the second-quantization Hamiltonian with the electron interaction into an effective Hamiltonian quadratic in creation and annihilation electron operators. The effective Hamiltonian   can be diagonalized by the Bogolyubov\,\textendash\,Valatin transformation.

\begin{figure}[b]%
\centering
\includegraphics[width=0.5\textwidth]{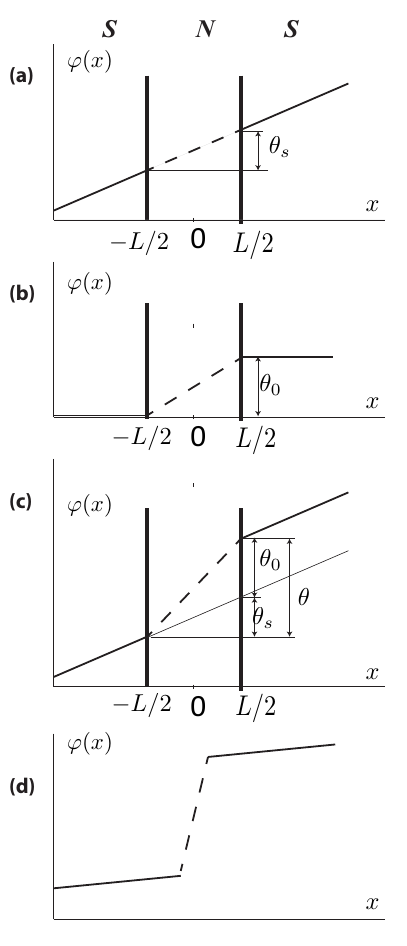}
\caption{The phase variation across the SNS junction.  (a) The condensate current produced by the phase gradient  $\nabla \varphi$ in the superconducting layers.  The phase $\theta_s =L\nabla \varphi$ is the superfluid phase. In all layers the electric current is equal to $env_s$.  (b) The vacuum current produced by the phase $\theta_0=\theta_+-\theta_-$, which is called the vacuum phase. The current is confined to the normal layer, there is no current in superconducting layers.  (c) The  superposition of the condensate  and the vacuum current. (d) The phase variation across a weak link.}\label{f1}
\end{figure}

The effective Hamiltonian is not gauge invariant, and the theory using this Hamiltonian  violates the charge conservation  law. The charge conservation law is restored  if one solves the Bogolyubov\,\textendash\,de Gennes equations  together with the self-consistency  equation for the pairing potential. In the past \cite{Kulik,Ishii,Bard} this step was skipped. Instead of solving the self-consistency  equation, it was  postulated that there is  a gap $\Delta$ of constant modulus $\Delta_0=\rvert \Delta\rvert $  in the superconducting layers and zero gap inside the normal layer.\footnote{The same pairing potential profile was assumed in the original paper by Andreev \cite{And65}.} Further we use this idealized model assuming the following spatial variation  of the gap in space (Fig.~\ref{f1}):
\be
\Delta =\left\{ \begin{array}{cc} \Delta_0e^{i\theta_++ i\nabla \varphi x} & x>L/2 \\  0 & -L/2<x<L/2\\ \Delta_0e^{i\theta_-+ i\nabla \varphi x} & x<-L/2  \end{array} \right. .
   \ee{prof}
The effective masses and Fermi energies are equal in  the superconductors and in the normal metal. 

Previous theoretical investigations of the ballistic SNS junction   have left some questions open:
\begin{enumerate}

\item Since the effective Hamiltonian is not gauge invariant, some solutions of the model violate the charge conservation law. They are mathematically correct, but are unphysical and must be filtered out. In many previous papers starting from the original ones   \cite{Kulik,Ishii,Bard} it was not done properly.

\item
There was no clarity about the respective roles of bound and continuum states in the current in the normal layer. Ishii \cite{Ishii} argued that both were important. This view was shared by a number of later publications (see Thuneberg \cite{Thun} and references therein). But a clear quantitative analysis of the issue was still needed. 
\item
The effect of parity of Andreev levels (odd vs. even number of states) was not investigated or even mentioned. The effect is possible in the 1D case.

\end{enumerate}

 Reference \cite{Son21} addressed these issues recently. The  analysis dealt directly with {\em  ab initio} analytical expressions for relevant currents via sums and integrals over all bound  and continuum states without introducing Green's function formalism for their calculation as was mostly done in the past starting from Refs. ~\cite{Kulik,Ishii}. Sums and integrals were calculated by expanding them in the inverse thickness $1/L$ of the normal layer. 
 
 In paper \cite{Son21},  a  remedy for the violation of the conservation law in previous investigations was proposed. The strict conservation law was replaced by  a softer condition that, at least, the total currents deep in all layers are the same. The condition  can  be satisfied only by taking into account three contributions to the total current $J=J_s+J_v+J_q$: (i) The current $J_s$ is induced by the phase gradient in the superconducting layers. It was called  the Cooper-pair condensate, or simply the  condensate current.  (ii) The current $J_v$, which can flow  in the normal layer even if the Cooper-pair condensate is at rest  and all states are empty. It  was called  the vacuum current.  (iii) The current $J_q$  induced by nonzero occupation of Andreev levels, i.e., by creation of quasiparticles. It was called the excitation current. Since the condensate motion produces the same current $J_s$   in all layers, while the  vacuum  and the  excitation currents  exist only in the normal layer, the charge conservation law requires that the  sum  of the  vacuum  and   the excitation currents $J_v+J_q$  vanishes.

The solution of the Bogolyubov\,\textendash\,de Gennes equations for the phase variation  shown  in Fig.~\ref{f1}(a)  gives the state with the only condensate current $J_s=env_s$ in all layers. Here $n$ is the electron density and $v_s ={\hbar \over 2m}\nabla \varphi$  is the superfluid velocity. This current does not differ from the current in a uniform superconductor since at Andreev scattering at interfaces between  layers the Bogolyubov\,\textendash\,de Gennes equations are Galilean invariant despite the translational invariance is broken \cite{Bard,Son21}. The phase difference across the normal layer is $\theta_s=\nabla \varphi L$. It was called the superfluid phase  \cite{Son21}. There is a state with the vacuum current $J_v$ flowing only in the normal layer [Fig.~\ref{f1}(b)], which  is determined by the vacuum phase $\theta_0=\theta_+-\theta_-$  [see \eq{prof}]. The  charge conservation law is violated if there is no excitation current compensating the vacuum current. Figure~\ref{f1}(c) shows the phase variation at the coexistence of the condensate and the vacuum current. The total phase difference across the normal layer is the Josephson phase $\theta=\theta_s+\theta_0$. The phase profiles in the normal layer are shown in Fig.~\ref{f1} by dashed lines. This phase is not determined because it is a phase of the order parameter $\Delta$, which vanishes  in the normal layer. Only the total phase difference across the normal layer appears in the Bogolyubov\,\textendash\,de Gennes equations. The dashed lines simply show what the phase gradient would be if the normal metal were replaced by a superconductor.

In the past it was common to ignore  the effect of the phase gradient in superconducting leads on the Josephson current.  By default, it was supposed that this is not an issue because gradients in the leads are very  small. This is true for a weak link, inside which the phase varies much faster than in the leads as shown in Fig.~\ref{f1}(d). At zero temperature the long SNS junction is not a weak link \cite{Son21}, and the phase gradient in the leads does affect the current in the normal layer. Thus, one should determine   currents in the normal and  superconducting layers  self-consistently.  The long ballistic SNS junction is a weak link only at high temperatures when the current  is exponentially small.

\begin{figure}[!b] 
\centering
\includegraphics[width=0.4 \textwidth]{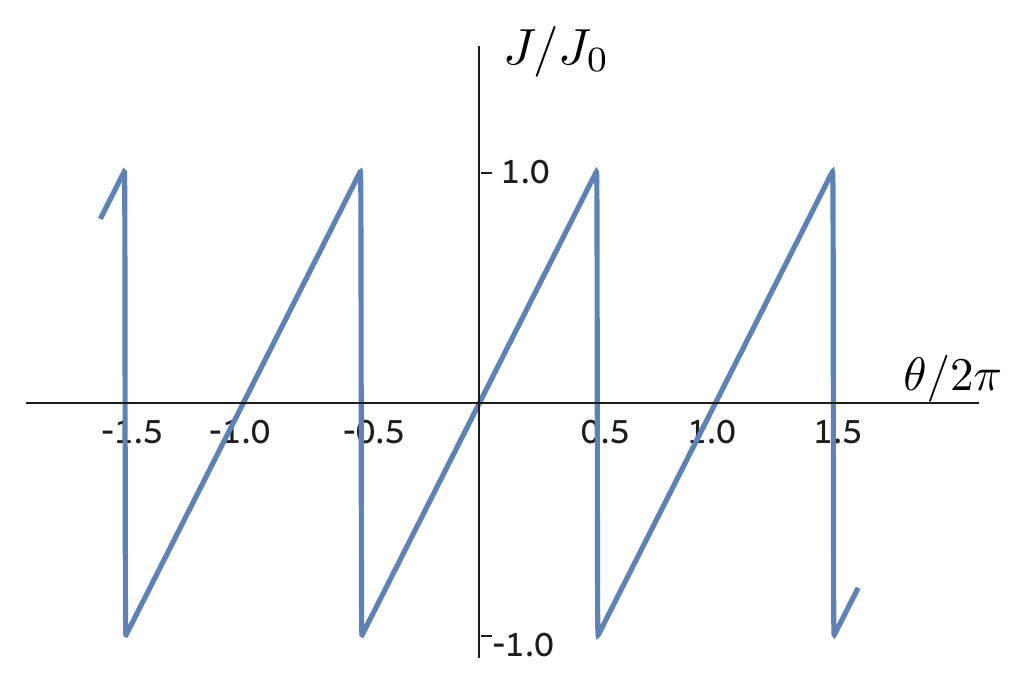} 
 \caption{The saw-tooth current-phase relation at zero temperature. Here $J_0={\pi\hbar\over 2mL}en$ ($={ev_f\over L}$ in the 1D case).\label{ST}}
 \end{figure}

Introduction of two phases $\theta_0$ and $\theta_s$ and the difference between the condensate and the vacuum current essentially revised the physical picture of charge transport through the ballistic SNS junction  at zero temperature. The principal difference between two phases is that at tuning the phase $\theta_0$ Andreev levels move with respect to the gap, while at tuning the phase $\theta_s$  Andreev levels move together with the gap and their respective positions do not vary. At zero temperature all investigations predicted  a saw-tooth current-phase relation (Fig.~\ref{ST}). But in the works    \cite{Kulik,Ishii,Bard} ignoring phase gradients in the leads, the current  at sloped segments was a vacuum current calculated using the formalism of finite temperature Green's functions. As argued above, the vacuum current cannot flow alone because this violates the charge conservation law. By contrast, according to Ref.~\cite{Son21}, at sloped segments only  the condensate current $J_s$ flows in all layers without violation of the charge conservation law. Its value directly follows from the Galilean transformation of the ground state in the condensate at rest to the state with moving condensate. As demonstrated in Sec.~\ref{Gal}, the Bogolyubov\,\textendash\,de Gennes equations are Galilean invariant if there is no normal scattering and the Andreev reflection is the only scattering process. The latter condition is satisfied   if the ratio $\Delta_0/\varepsilon_f$ ($\varepsilon_f$ is the Fermi energy) is small. Thus, the derivation of the zero-temperature saw-tooth current-phase relation goes beyond the simple  model  of the step-function gap profile  (Fig.~\ref{ST}) and is valid even if the gap profile is determined from the self-consistency equation. For derivation it is not necessary to first use the sophisticated theory for finite temperature Green functions in order to go to the limit $T\to 0$ in the end.

The vacuum current at zero temperature appears only at the vertical segments of the $T=0$ current-phase relation at $\theta =\pi (2s+1)$ ($s$ is an integer) when  the energy of the  lowest   Andreev state reaches zero and its  occupation becomes  possible. This allows to satisfy the condition $J_v+J_q=0$.

The difference  between the phases $\theta_0$ and $\theta_s$ was already considered in the past by Riedel {\em et al.} \cite{Bagwell} although using a different terminology.\footnote{I am thankful to the anonymous referee  who attracted my attention  to this paper, which was unknown to me at writing \cite{Son21}. The phases $\theta_s$ and $\theta_0$ in the present paper and in \cite{Son21} correspond to the phases $\phi_R-\phi_L$ and $qL$ in \cite{Bagwell} respectively.}  They also realized an important difference between the point contact and  the planar long ballistic SNS junction: the former is a weak link and the latter is not. This was illustrated by their Fig.~1, which is equivalent to our Figs.~\ref{f1}(a) and \ref{f1}(d). The problem of charge conservation when the junction is not a weak link was discussed by Sols and Ferrer \cite{Sols} for a superconductor interrupted by a barrier with high transmission probability.

Riedel {\em et al.} \cite{Bagwell} went beyond the model with the postulated step-function dependence of the gap on the coordinate (Fig.~\ref{f1}) and numerically solved  the Bogolyubov\,\textendash\,de Gennes equations together with the integral self-consistency equation for the gap. Remarkably, they revealed that even though their numerics gave smoothly varying gap dependence in the transient interface between  the normal layer and the superconducting leads, the phase gradient remained constant along the whole junction as in Fig.~\ref{f1}(a). This is a numerical confirmation of the Galilean invariance proved in  Sec.~\ref{Gal} of the present paper analytically. Recently Davydova {\em et al.} \cite{diode} also discussed  the Galilean transformation  (Doppler shift) for a short SNS junction shunted by a  nanowire bridge.  This setup was suggested for observation of the Josephson diode effect and essentially differs from that in the present paper. In their case the state with  the only condensate current in all layers is absent.

Thuneberg published  the Comment  \cite{Thun} rejecting the approach in Ref.~\cite{Son21} and asserting that the vacuum current  in continuum states  was incorrectly ignored there (see the Reply \cite{Son23rep} to the Comment). According to \cite{Son21}, in multidimensional (2D and 3D) cases  the vacuum current in continuum states was   suppressed after integrating over wave vectors  transverse to the current through the junction. However, in the 1D case transverse degrees of freedom are absent, and the contribution of  continuum states to  the vacuum current does not vanish, but it was crudely estimated as insignificant  in the limit $L\to \infty$. In the aftermath of the discussion with Thuneberg this estimation for the 1D case was re-accessed. It was revealed that the estimation missed an important term of the same order $\propto 1/L$  as the current  in bound states.    A more careful  calculation of the vacuum current  in continuum states  is presented in this paper. Thus, Thuneberg  \cite{Thun} was vindicated in the 1D case. But the respective roles of bound and continuum states are different  in the 1D and the multidimensional cases  \cite{Son21}.

Discussing the controversy about  respective contributions of bound and continuum states into the current, we have in mind only the vacuum current. As for the condensate current, it is clear that contributions of bound and continuum states into the current are proportional to their contributions to the density. The contribution of bound states to the density is  less than  the contribution of continuum states by the small factor $\Delta_0/\varepsilon_f$.

The re-evaluation of the contribution of continuum states to the vacuum current refuted the prediction \cite{Son21}  that in the 1D case the ballistic long SNS junction becomes a $\varphi_0$ junction at high temperatures. The $\varphi_0$ junction is an anomalous Josephson junction,  in which the ground state is not at zero phase.\footnote{In Ref.~\cite{Son21} they were called the $\theta$ junctions, but the term $\varphi_0$ junction is more common in the literature \cite{Buzdin}.}  The prediction of a $\varphi_0$ junction was based on the existence of  the large $\propto  1/L$ temperature-independent term in the excitation current with the sign opposite to the sign of the phase, which was not compensated by the vacuum current in bound states. However, according to the present analysis, the large negative excitation current is fully compensated by the positive continuum vacuum current, which was not taken into account in Ref.~\cite{Son21}.

In Ref.~\cite{Son21} the effect  of incommensurability of the gap magnitude in the superconducting layers with respect  to the Andreev level energy spacing  was investigated in the 1D case.  With tuning of the incommensurability parameter, the  number of bound Andreev states varies being odd or even for every spin (the parity effect). This produces jumps on the current-phase relation.  However, the present analysis  reveals that the parity effect exists also in the continuum states, which  were ignored in Ref.~\cite{Son21}. Although there is no discrete states at energies exceeding the gap, there are peaks of the transmission probability for electrons  crossing the normal layer (transmission resonances). The energy spacing between peaks is the same as the energy spacing between Andreev levels.  While tuning phase, a new Andreev level appears (or an old one disappears) inside the gap, correspondingly some peak  disappears (or a new one appears) in the continuum. As a result, the two parity effects compensate each other, and the total effect vanishes. Transmission peaks were known {in the past \cite{Bard,Svidz71,Kummel}. } Although the parity effect is cancelled in the total vacuum current, the effect yields important contributions $\propto 1/L$ to  the currents in the bound and continuum states calculated separately.

The analysis in the paper is focused on the 1D case, where the only motion is along the axis $x$ normal to layers. For the extension of the calculation  on multidimensional systems
currents must be integrated over  transverse components   of   wave vectors.

\section{The Bogolyubov\,\textendash\,de Gennes theory}
 \label{BdG}

 In the  self-consistent field method  \cite{deGen} the  second-quantized effective Hamiltonian density quadratic in the wave function is
\bem
{\cal H}_{eff}={\hbar^2 \over 2m} [\nabla \hat\psi^\g_\gamma(x) \nabla \hat\psi_\gamma(x)- k_f^2 \hat\psi^\g_\gamma(x) \hat\psi_\gamma(x)] 
\nonumber \\
+\Delta \hat\psi^\g_\ua(x) \hat\psi^\g_\da(x)+\Delta^* \hat\psi_\da(x) \hat\psi_\ua(x), 
  \eem{ef}
where $\hat\psi_\gamma^\g(x)$ and $\hat\psi_\gamma(x)$ are operators of creation and annihilation of an electron, and the subscript $\gamma$ has two values corresponding to the spin up ($\ua$) and  down ($\da$).
We address a 1D problem  with the Fermi  wave number $k_f$, assuming that our system is uniform in the plane normal to the axis $x$.  In multidimensional systems with the Fermi wave number $k_F$   $k_f =\sqrt{k_F^2 -k_\perp^2}$, where $k_\perp$ is the transverse component of the multidimensional wave vector $\bm k$. 

The effective Hamiltonian can be diagonalized by the Bogolyubov\,\textendash\,Valatin transformation from the free electron operators $\hat\psi_\gamma^\g(x)$ and $\hat\psi_\gamma(x)$ to  the quasiparticle operators $\hat a_{i\gamma}^\g$ and $\hat a_{i\gamma}$:
\bem
\hat \psi_\ua(x) =\sum_{i} \left[u_i(x) \hat a_{i\ua} -v^*_i(x) \hat a^\dagger_{i\da}   \right],
\nonumber \\
\hat \psi_\da(x) =\sum_{i} \left[u_i(x) \hat a_{i\da}+v^*_i(x) \hat a^\dagger_{i\ua}   \right].
    \eem{BV}
The summation over the subscript $i$ means the summation over all bound and continuum states. The functions $ u(x,t)$ and $v(x,t)$ are two components of  a spinor wave function,
\begin{equation}
\psi(x,t) = \left[ \begin{array}{c} u(x,t)\\  v(x,t) \end{array}
\right], 
     \label{spinor} \end{equation}
describing a state of a quasiparticle, which is a superposition of a state with one particle (upper component~$u$) and a state with
one antiparticle, or hole (lower component~$v$). They are stationary solutions of 
the Bogolyubov\,\textendash\,de Gennes equations:
\bem
i\hbar {\partial u\over \partial  t} ={\delta{\cal H}_{BG}\over \delta u^*} =-{\hbar^2 \over 2m} \left( \nabla^2 + k_f^2\right) u  + \Delta  v ,
\nonumber \\
i\hbar {\partial v\over \partial  t}  ={\delta{\cal H}_{BG}\over \delta v^*}={\hbar^2 \over 2m}  \left( \nabla ^2 + k_f^2\right) v  + \Delta^* u .
     \eem{BG}
 The Bogolyubov\,\textendash\,de Gennes equations are the  Hamilton equations with  the Hamiltonian (per unit volume) 
\bem
{\cal H}_{BG}={\hbar^2 \over 2m} (\lvert \nabla u\rvert^2 - k_f^2  \lvert u\rvert^2)-{\hbar^2 \over 2m} ( \lvert \nabla v\rvert^2 - k_f^2  \lvert v\rvert^2)
+\Delta u^* v+\Delta^*v^* u.
   \eem{ham}

  The number of particles (charge) is not a quantum number of the state.  The average  density $n_i$ and current $j_i$ in the $i$th state are
\be
n_i= \lvert u_i\rvert^2-\lvert v_i\rvert ^2, 
   \ee{ni}
and
\be
j_i= -{ie\hbar\over 2m} (u_i^*\nabla u_i-u_i\nabla u_i^*) -{ie\hbar\over 2m} (v_i^*\nabla v_i-v_i\nabla v_i^*). 
 \ee{ji} 

After the diagonalization the effective  Hamiltonian becomes
\be
{\cal H}_{eff}=\sum_i\varepsilon_i(\hat a^\g_{i\ua} \hat a_{i\ua}+\hat a^\g_{i\da} \hat a_{i\da}-2\lvert v\rvert^2),
  \ee{efd}
where $\varepsilon_i$ is the energy of the $i$th quasiparticle state.  
 
  The total density $n$  and the total charge  current $J$ are  expectation  values for the   operators 
\bem
\hat n(x)=\hat \psi^\dagger_\ua(x)\hat \psi_\ua(x)+\hat \psi^\dagger_\da(x)\hat \psi_\da(x)
 \nonumber \\ 
=\sum_{i}\left[ \lvert u_i(x)\rvert ^2  (\hat a^\dagger_{i\ua} \hat a_{i\ua}+  \hat a^\dagger_{i\da} \hat a _{i\da})+\lvert v_i(x)\rvert ^2(\hat a_{i\ua} \hat a^\dagger_{i\ua}+  \hat a_{i\da} \hat a^\dagger _{i\da} )\right]
\nonumber \\ 
=\sum_{i}\left\{ [\lvert u_i(x)\rvert ^2  -\lvert v_i(x)\rvert ^2 ](\hat a^\dagger_{i\ua} \hat a_{i\ua}+\hat a^\dagger _{i\da} \hat a_{i\da}) +2 \lvert v_i(x)\rvert ^2\right\},
  \eem{nOp}
\bem
\hat J= -{ie\hbar\over 2m}\sum_{i} \left[(u_i^*\nabla u_i-u_i\nabla u_i^*+v_i^*\nabla v_i-v_i\nabla v_i^*)(\hat a^\dagger_{i\ua} \hat a_{i\ua} +  \hat a^\dagger _{i\da} \hat a_{i\da})
\right. \nonumber \\ \left.
-2 (v_i^*\nabla v_i-v_i\nabla v_i^*)\right].
 \eem{n-j}

There are two contributions to the   energy,  the density, and the current [Eqs.~(\ref{efd})--(\ref{n-j})]. One  is connected with the quasiparticle vacuum, in which all energy levels are  not occupied (last terms in  equations without operators). This contribution is responsible for the condensate and the vacuum current. The other terms in   the  equations  are connected with quasiparticles  occupying  energy levels. They are responsible for the excitation current.

In a uniform superconductor at rest, the constant $\Delta_0$ solutions of the Bogolyubov\,\textendash\,de Gennes equations are plane waves
\begin{equation}
\left(  \begin{array}{c}
u_0    \\
  v_0 
\end{array}  \right)e^{i k \cdot x-i\varepsilon_0  t/\hbar},
   \label{pwsol}
\end{equation}
where
\be
u_0=\sqrt{{1\over 2} \left( 1+  {\xi\over \varepsilon_0 }\right)} ,~~v_0= \sqrt{{1\over 2} \left( 1-  {\xi\over \varepsilon_0 }\right)}.
          \ee{u0v0}
The quasiparticle energy is given by the well known BCS expression: 
\begin{equation}
\varepsilon_0 =  \sqrt{\xi^2 + \Delta_0^2}.
     \label{SpBCS}
\end{equation}
 Here  $\xi = ({\hbar^2 / 2m})(k^2 - k_f^2)\approx \hbar v_f (k-k_f)$  is the quasiparticle energy in the normal Fermi liquid, and $v_f=\hbar  k_f/m$ is the  Fermi velocity.  
The states with positive and the negative signs of $\xi$ correspond to particle-like and the hole-like  branches of the spectrum respectively. 

The Bogolyubov\,\textendash\,de Gennes equations have also solutions with negative  energies $\varepsilon_0=  - \sqrt{\xi^2 + \Delta_0^2}$. In a stable state, energies of all excitations must be positive, and solutions with  negative energies should not be considered   \cite{Tin}.
 
\section{{\em Ab initio} expressions for currents }
 \label{ASSNS}

\subsection{Galilean invariance and condensate current } \label{Gal}

An important assumption in the present analysis (as well as in the previous investigations) was that only the Andreev reflection is possible on interfaces between superconducting  
and normal layers. The assumption is valid in the limit of large Fermi wave numbers $k_f \gg \Delta_0 /\hbar v_f$.  As mentioned in Introduction, this means that there is no significant change of the quasiparticle momentum after reflection, and  
wave functions $\left(  \begin{array}{c}
u    \\
  v 
\end{array}  \right)$  are superpositions of plane  waves with wave numbers only close to either  $+ k_f$, or  $- k_f$. These plane waves describe quasiparticles, which  will be  called  rightmovers (+) and leftmovers (-).  After transformation of the wave function,
\be
\left(  \begin{array}{c}
u    \\
  v 
\end{array}  \right) =\left(  \begin{array}{c}
\tilde u    \\
\tilde  v 
\end{array}  \right)e^{\pm ik_f x},
        \ee{}
the second order  terms   in gradients, $\nabla^2 \tilde  u$ and $\nabla^2 \tilde  v$, can be neglected for small $\Delta_0/\varepsilon_f$, and the Bogolyubov\,\textendash\,de Gennes equations are reduced to the equations of the first order  in gradients:
\bem
\varepsilon \tilde u  =\mp i\hbar v_f \nabla \tilde  u  + \Delta \tilde  v ,
\nonumber \\
\varepsilon \tilde v =\pm i\hbar v_f \nabla \tilde v  + \Delta^*\tilde  u .
    \eem{BG1}
The boundary conditions on the interfaces between layers require the continuity  of the wave function components, but not their gradients.

Let us demonstrate Galilean invariance of the Bogolyubov\,\textendash\,de Gennes equations when the wave functions are superpositions of only rightmovers, or only of leftmovers. Suppose that we found the Bogolyubov\,\textendash\,de Gennes function    $\left(  \begin{array}{c}
\tilde u_0    \\
\tilde  v_0 
\end{array}  \right)$ with the energy $\varepsilon_0$
 for an arbitrary profile of the superconducting gap $\Delta(x)$. Now we check what is the solution of the Bogolyubov\,\textendash\,de Gennes equations for the superconducting gap $\Delta(x)e^{i\nabla\varphi x}$, where $\nabla\varphi$ is a constant phase gradient.
 The wave function 
 \be
 \left(  \begin{array}{c}
\tilde u    \\
\tilde  v
\end{array}  \right)=\left(  \begin{array}{c}
\tilde u_0 e^{i\nabla\varphi x/2}   \\
\tilde  v_0  e^{-i\nabla\varphi x/2} 
\end{array}  \right)
        \ee{} 
satisfies \eq{BG1} where the gap $\Delta(x)$ is replaced by    $\Delta(x)e^{i\nabla\varphi x}$ and the energy  is 
\be
 \varepsilon=\varepsilon_0 \pm   {\hbar v_f\over 2}\nabla\varphi=\varepsilon_0 \pm   {\hbar v_f\over 2L}\theta_s=\varepsilon_0 \pm   v_s \hbar k_f.
      \ee{GT}   
Thus, the Galilean transformation produces the same Doppler shift in the energy as in a uniform superconductor. According to Eqs.~(\ref{ni}) and (\ref{ji}) for the density $n_i$  and the current $j_i$ in the $i$th Bogolyubov\,\textendash\,de Gennes state, the Galilean transformation transforms the current $j_i$ to the current $j_i+en_i v_s$.  Summation over all bound and continuum states  yields that the Galilean transformation added the  condensate current $J_s=env_s$ in the whole space. The difference between the  total charge densities $n$ in the normal and the superconducting  layers vanishes in the limit $L\to \infty$ \cite{Son21}. Thus, the condensate current is the same in all layers and does not violate the charge conservation law. 

Our derivation does not depend on the profile of the gap in the space, as far as the gap is small compared to the Fermi energy and there is no normal scattering transforming rightmovers into leftmovers and {\em vice versa}.  The Doppler shift of the energy is of opposite sign for rightmovers and leftmovers, and after the Galilean transformation of a superposition of rightmovers and leftmovers the state is not an eigenstate of the energy. The derivation remains valid if $\Delta$ vanishes in some part of the space. Independence of the condensate current on the gap profile is confirmed by numerical solution of  the Bogolyubov\,\textendash\,de Gennes equations together with the integral self-consistency equation for the gap by Riedel {\em et al} \cite{Bagwell}, as already mentioned in Sec.~\ref{Intr}. They obtained that although the absolute value of the gap smoothly varies between the normal layer and the  superconducting leads, the phase gradient is constant in all layers as in Fig.~\ref{f1}(a). Summarizing, the value of the condensate current directly follows from the Galilean invariance and does not need sophisticated calculations.

\subsection{Vacuum current in bound states}

The spectrum and  the wave functions for the present model of the SNS junction are well known from previous works, and it is sufficient here to  present the resume of these investigations. 
Solving the Bogolyubov\,\textendash\,de Gennes equations with the  gap profile given by \eq{prof} for zero gradient $\nabla \varphi$ (Cooper-pair condensate at rest) and  with the boundary conditions formulated above one finds the equation for  energies  $\varepsilon_{0\pm} (s)$ of  bound Andreev states:
\be
\varepsilon_{0\pm} (s)={\hbar v_f\over 2L}\left[2\pi  s+2 \arcsin {\sqrt{\Delta_0^2-\varepsilon_{0\pm} (s)^2}\over \Delta_0} \pm \theta_0\right].
   \ee{eps00}
Here  $s$ is an  integer. The upper and the  lower signs correspond to   rightmovers (+) and leftmovers (-) respectively. If the condensate moves ($\nabla \varphi \neq 0$) the energy $\varepsilon_\pm (s)$ of the state is given by \eq{GT}:
\be
\varepsilon_\pm (s)={\hbar v_f\over 2L}\left[2\pi  s+2\arcsin {\sqrt{\Delta_0^2-\varepsilon_{0\pm} (s)^2}\over \Delta_0} 
 \pm \theta\right].
   \ee{eps0}
The phase in the equation for  $\varepsilon_{0\pm} (s)$ is $\theta_0$, while the energy $\varepsilon_\pm(s)$ depends on the total Josephson phase  $\theta=\theta_0 +\nabla \varphi L=\theta_0 +\theta_s$.

At small energy $\varepsilon_{0\pm} (s) \ll \Delta_0$ (small $s$) 
\be
\varepsilon_{0\pm} (s)={\hbar v_f\over 2L}\left[2\pi  \left(s+{1\over 2}\right)\pm \theta_0\right].
   \ee{en0}

For  Andreev levels close to the gap one can  expand the arcsin function in \eq{eps00} transforming it to 
\be
 \Delta_0- \varepsilon_{0\pm}   -{\zeta_0\over L} \sqrt{2\Delta_0(\Delta_0- \varepsilon_{0\pm} )}=\Delta_0-\pi  s{\zeta_0 \over L}\Delta_0=\pi ( t+\alpha){\zeta_0 \over L}\Delta_0.
 \ee{eqD}

Here  $\alpha$ ($0 <\alpha <1 $)  is the parameter of incommensurability, which is the fractional part of the ratio of the gap $\Delta_0$ to the Andreev level energy spacing,
\be
{\Delta_0  L\over \pi \hbar v_f}={L\over \pi\zeta_0}=s_m +\alpha,
     \ee{inc}
 $s_m$ is the maximal integer less than the ratio, $t=s_m-s$ is another integer, and 
\be
\zeta_0={\hbar v_f\over  \Delta_0} 
     \ee{calL}
is the coherence length. Solution of \eq{eqD}  yields
\bem
\varepsilon_{0\pm} =\Delta_0-{\pi \zeta_0\over L}\Delta_0\left[\sqrt{t +\alpha  \mp  {\theta_0\over  2\pi }+{\zeta_0\over 2\pi L }}-\sqrt{{\zeta_0\over 2\pi L}}\right]^2,
\nonumber \\
\sqrt{\Delta_0^2- \varepsilon_{0\pm} ^2}=\Delta_0\sqrt{2\pi \zeta_0 \over L}\left[\sqrt{t +\alpha  \mp  {\theta_0\over  2\pi }+{\zeta_0\over 2\pi L }}
-\sqrt{\zeta_0\over 2\pi L}\right].
    \eem{Edel}

There is an essential difference between effects of phases $\theta_0$ and $\theta_s$ on the Andreev spectrum \cite{Son21}. Variation of $\theta_0$ makes the Andreev levels to move with respect  to the gap. As a result, some new levels can emerge and some old ones can disappear. In contrast, variation of $\theta_s$ leads  to the shift of the spectrum as a whole without changing positions of Andreev levels with respect to the gap.  The principle  of the BCS theory that only solutions with positive energies should be taking into account refers to the energy  $\varepsilon_0$, while the Doppler-shifted  energy  $\varepsilon$ can be both positive or negative. If   $\varepsilon$ is negative the level is occupied at zero temperature. 

 The current in the Andreev state is determined by the canonical relation connecting it with the derivative of the  energy with respect to the phase:
\be
j_\pm(s)={2e\over \hbar}{\partial  \varepsilon_{0\pm} (s)\over \partial \theta_0}=\pm{e\over \pi \hbar}{\partial  \varepsilon_{0\pm} (s)\over \partial s}=\pm{ev_f\over L+\zeta}
%\nonumber \\
=\pm {ev_f\over L}\frac{\sqrt{1- {\varepsilon_{0\pm} (s)^2\over \Delta_0^2}}}{\sqrt{1- {\varepsilon_{0\pm} (s)^2\over \Delta_0^2}}+{\zeta_0\over L}}.
       \ee{js}
   Here 
   \be
\zeta= \zeta_0{\Delta_0\over \sqrt{\Delta_0^2-\varepsilon_0^2}}
  \ee{zeta}
is   the depth of penetration of the bound states into the superconducting layers, which diverges when  $\varepsilon_0$ approaches to the gap $\Delta_0$. The factor 2 takes into account that $\theta_0$ is the phase of a Cooper pair but not of a single electron.

The current $j_\pm(s)$ is a current produced by a quasiparticle created at the $s$th state. However, we look for the vacuum current when there is no quasiparticles. In Andreev states $\vert u\rvert^2=\vert v\rvert^2={1\over 2}$, and according to \eq{n-j} the vacuum current in any state is two times less, and it has a  sign opposite to the sign of the  current $j_\pm(s)$. Taking this into account and including two spin states, the {\em ab initio} expression for the vacuum current in  bound states is 
\bem
J_{vA}=-\sum_s \{j _+(s)\mbox{H}[\varepsilon_{0+} (s)]  \mbox{H}[\Delta_0 -\varepsilon_{0+} (s)] +j_-(s)\mbox{H}[\varepsilon_{0-} (s)]  \mbox{H}[\Delta_0 -\varepsilon_{0-} (s)]\}
\nonumber \\
=-{ev_f\over L}\sum_s \left\{\frac{ \sqrt{1- {\varepsilon_{0+} (s)^2\over \Delta_0^2}} }{\sqrt{1- {\varepsilon_{0+} (s)^2\over \Delta_0^2}} +{\zeta_0\over L}}\mbox{H}[\varepsilon_{0+} (s)]  \mbox{H}[\Delta_0 -\varepsilon_{0+} (s)]
\right. \nonumber \\  \left.
- \frac{ \sqrt{1- {\varepsilon_{0-} (s)^2\over \Delta_0^2}} }{\sqrt{1- {\varepsilon_{0-} (s)^2\over \Delta_0^2}} +{\zeta_0\over L}}\mbox{H}[\varepsilon_{0-} (s)]  \mbox{H}[\Delta_0 -\varepsilon_{0-}(s)]\right\} .~~~
           \eem{Jb}
Here  $\mbox{H}(q)$ is the Heaviside step function, which ensures that summation over $s$ extends only on states with energies $0<\varepsilon_{0\pm} <\Delta_0$ inside the gap. 

\subsection{Continuum vacuum current}

According to  Refs.~\cite{Bard,Son21,Bagw},    for  a rightmover quasiparticle  ($\xi>0$) incident from left  the transmission   and the reflection probabilities are 
\begin{equation}
{\cal T}(\theta_0)= \frac{2(\varepsilon_0 ^2-\Delta_0^2)} {2\varepsilon_0^2 -\Delta_0^2-\Delta_0^2\cos\left({2\varepsilon_0 m  L\over \hbar^2 k_f}-\theta_0\right)},~~R(\theta_0)=1-{\cal T}(\theta_0).
\label{T} \end{equation}
These expressions are valid also for a quasihole ($\xi<0$) incident from right. For a quasiparticle incident from right and a hole incident from left   the transmission   and the reflection probabilities are ${\cal T}(-\theta_0)$ and $R(-\theta_0)$.

\Eq{T} was obtained in the coordinate frame moving with the condensate. Transformation to  the laboratory frame leads to replacing of the argument ${2\varepsilon_0 m  L\over \hbar^2 k_f}-\theta_0$ of the cosine function by  ${2\varepsilon m  L\over \hbar^2 k_f}-\theta$. According to   \eq{GT} they are equal. Thus, the Galilean transformation discussed in Sec.~\ref{Gal} does not change the scattering parameters. 

One can transform expressions for ${\cal T}$ and $R$ revealing their  dependence on the incommensurability parameter $\alpha$ introduced in \eq{inc}:
\begin{equation}
{\cal T}(\theta_0)= \frac{2(\varepsilon_0 ^2-\Delta_0^2)} {2\varepsilon_0 ^2 -\Delta_0^2-\Delta_0^2\cos\left[{2(\varepsilon_0 -\Delta_0)m  L\over \hbar^2 k_f}+2\pi \alpha-\theta_0\right]}.
     \label{Ta} \end{equation}
The reflection  probability can be transformed similarly. The both probabilities rapidly oscillate  as functions of the energy.

Collecting together all contributions from rightmovers and leftmovers, quasiparticles and quasiholes, the  {\em ab initio} expression for  the continuum vacuum  current is \cite{Son21}
\be
J_{vC} ={e\over \pi\hbar}\int_{\Delta_0}^\infty  [{\cal T}(- \theta_0)-{\cal T}( \theta_0) ] d\xi.  
     \ee{cJ}

\subsection{Excitation current}

 As well as in  previous literature, the present analysis addresses the case when temperatures are much lower than critical  ($T \ll \Delta_0$). Then one can ignore excitations in continuum states and replace sums for a large but finite number of Andreev states by infinite sums. The {\em ab initio} expression for the  excitation current assumes the Fermi distribution  in Andreev levels:
\bem
J_q=  \sum_{s=0}^\infty\frac{2j_+(s)}{e^{\varepsilon_+(s)/T}+1}+\sum_{s=0}^\infty\frac{2j_-(s)}{e^{\varepsilon_-(s)/T}+1}.
      \eem{JqS}
The currents $j_\pm(s)$ in the $s$th state are given by \eq{js} derived for the condensate at rest despite this expression is  for the case of a moving condensate. This is because 
 in Andreev states $\lvert u \rvert^2=\lvert v \rvert^2$, and  according to \eq{ni} quasiparticle creation in an Andreev state does not change the electron density. Thus, the Galilean transformation does not change the electron current of the quasiparticle.  At $T \ll \Delta_0$ the expression \eq{en0} for the small energy $\varepsilon_{0\pm}$ can be used, and taking into account the relation \eq{GT}    between $\varepsilon_\pm(s)$ and $\varepsilon_{0\pm}(s)$, the excitation current is
\bem
J_q={2ev_f\over L} \sum\limits_0^\infty \left[\frac{1}{e^{\beta \left(s+{\pi+\theta \over 2\pi}\right)}+1}-\frac{1}{e^{\beta \left(s+{\pi-\theta \over 2\pi}\right)}+1}\right],
    \eem{JqIn} 
where
\be 
\beta={\pi \hbar v_f\over LT}.
  \ee{}
 While the vacuum current depends from the vacuum phase $\theta_0$, the excitation current depends on the total Josephson phase $\theta=\theta_0+\theta_s$.
 
\section{Calculation of vacuum and excitation currents}\label{calc}

Through the whole paper we assume in calculations that $0<\alpha<1/2$ and $\theta_0<\pi$. If  $1/2<\alpha<1$ currents of rightmovers and leftmovers will be different, but their sum will be the same  as for $0<1-\alpha<1/2$. As for the dependence  on $\theta_0$, its extension on the infinite interval of $\theta_0$ is straightforward bearing in mind that the current is an odd periodic function of $\theta_0$.

\subsection{Vacuum current  in bound Andreev states}

Calculating the vacuum current  in bound Andreev states one can expand the current in $1/L$. In Ref.~\cite{Son21} the first  terms $J_1\propto 1/ L$ and $J_{3/2}\propto 1/ L^{3/2}$ were calculated: 
\be 
J_{vA}= J_1+J_{3/2}.
      \ee{}
where \be
J_1     ={ev_f\over L}\mbox{H}(- \gamma_+ ),
     \ee{}
and
\bem
J_{3/2}={e\Delta_0\zeta_0^{3/2}\over \sqrt{2\pi} \hbar L^{3/2}}\left[\sum_{t=0}^{\infty} \left(\frac{1}{\sqrt{t+{\gamma_+\over  2\pi }  } }
- \frac{1}{\sqrt{t+{\gamma_-\over  2\pi }  } }\right)-\sqrt{2\pi \over \gamma_+ } \mbox{H}\left(-\gamma_+\right)  \right]
\nonumber \\
={e\Delta_0\zeta_0^{3/2}\over \sqrt{2\pi} \hbar L^{3/2}}\left[\zeta\left({1\over 2},{\gamma_+\over 2\pi}\right)
- \zeta\left({1\over 2},{\gamma_-\over  2\pi }\right)-\sqrt{2\pi \over \gamma_+ }\mbox{H}\left(-\gamma_+\right)  \right].
  \eem{Jsum}
Here
\be
\gamma_\pm= 2\pi \alpha\mp \theta_0,
     \ee{}
and\be
\zeta\left(z,q\right)=\sum_{t=0} ^\infty{1\over (q+t)^z}
      \ee{}
is Riemann's zeta function \cite{5}. The series for Riemann's zeta function at $z=1/2$ diverges, but the series for a difference of zeta functions with different arguments $q$  converges at large  $t$.
It was taken into account that the main contribution to the sum in \eq{Jsum} is given by Andreev levels close the gap, where one can  use the expression \eq{Edel} for $\varepsilon_{0\pm} (s)$.

At small positive $\gamma_+ =2\pi \alpha-\theta_0 \ll 1$ (but still $\gamma_+ \gg \zeta_0/L$) the current $J_{3/2}$ is determined only by the   term $t=0$ divergent in the limit $\gamma_+ \to 0$:
\be
J_{3/2}={e\Delta_0\zeta_0^{3/2}\over \hbar L^{3/2}\sqrt{\gamma_+}}.
      \ee{jAs}
Later we shall see that the current $J_{3/2}$ in the bound states is compensated by the $\propto 1/L^{3/2} $ current in continuum states.  

\subsection{Continuum vacuum current}

The continuum vacuum current is a difference  of contributions from right- and leftmovers:
\be
J_{vC} =J_+-J_- ,~~  J_\pm= - {e\over \pi\hbar}\int_{\Delta_0}^\infty  {\cal T}(\pm \theta_0) d\xi.  
     \ee{}
 Calculating $ J_\pm$ we introduce a new variable $z=(\varepsilon_0-\Delta_0)/\Delta_0$:
\bem
J_\pm=  -{e\Delta_0\over \pi\hbar}\int_0^{x_m}\frac{2\sqrt{2z+z^2}(1+z)\,dz}{4z+2z^2+1-\cos(2Lz/\zeta_0+ \gamma_\pm)}. 
   \eem{jvc}
The integrals for $J_\pm$ diverge, but their difference does not. So,  we introduced a  large $x_m$ as an upper limit of integrals assuming that in the end   $x_m\to \infty$.

\begin{figure}[h]%
\centering
\includegraphics[width=0.5\textwidth]{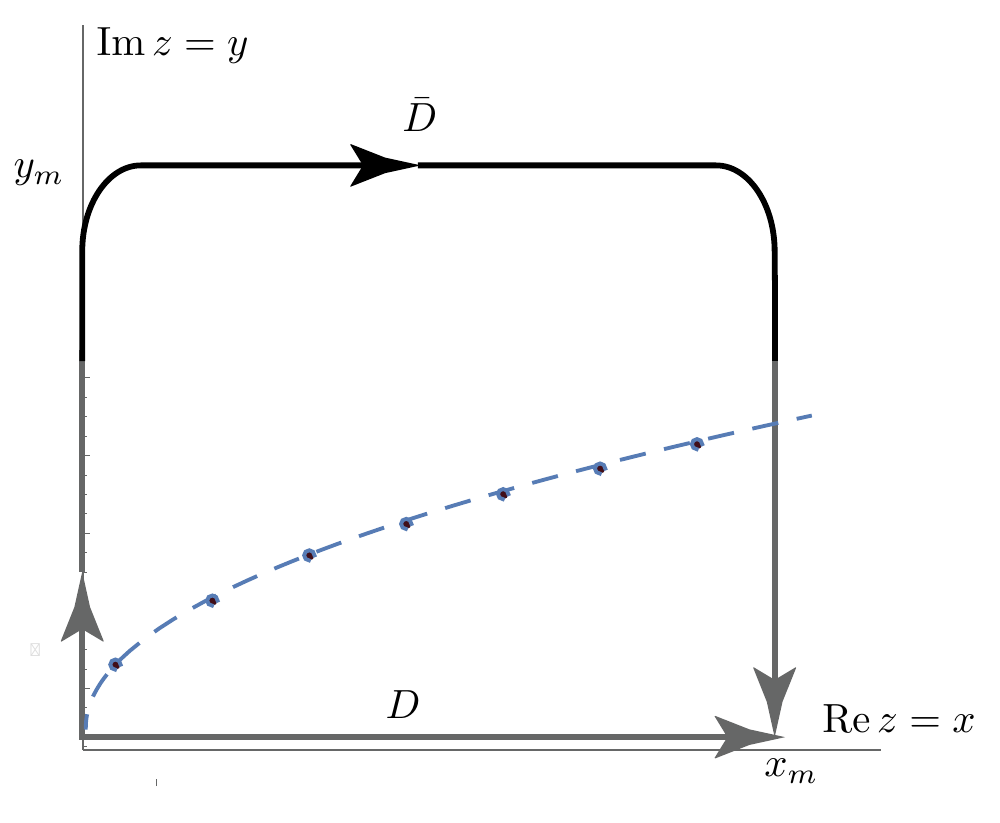}
\caption{Analytical continuation  in the complex plane $z=x+iy$ of the path for the integration in \eq{jvc}. Black circles on the dashed line are poles.} \label{CP}
\end{figure}

For calculation of the integrals  we  perform an analytical continuation from the real axis (contour $D$ in Fig.~\ref{CP}) to the complex plane $z=x+iy$ (contour $\bar D$ in Fig.~\ref{CP}). Calculating contributions of various segments of the contour we need to take into account only their real parts. The final integral \eq{jvc} is real, and the total sum of imaginary parts must vanish.

At the continuation the contour crosses   poles in points $z_{p\pm}=x_{p\pm}+iy_{p\pm}$. The values of $z_{p\pm}$ are roots of the equation
\be
1+4z_{p\pm}+2z_{p\pm}^2-\cos \left(2Lz_{p\pm}/\zeta_0+\gamma_\pm \right)=0,
    \ee{}
where $p$ is an integer number of a pole. At $L\to \infty$ the coordinates of poles  are
\be
x_{p\pm} ={\zeta_0\over 2L}(2\pi p -\gamma_\pm),~~y_{p\pm}={\zeta_0\over L}\ln(\sqrt{2x_{p\pm}+x_{p\pm}^2}+1+x _{p\pm} ).
    \ee{poles}
Residues of poles are
\bem
{\cal R}_{p\pm}=-{e\Delta_0\over \pi\hbar}\frac{2 \sqrt{z_{p\pm}(2+z_{p\pm})}(1+z_{p\pm})}{4+4z_{p\pm}+{2L\over \zeta_0}\sin \left(2Lz_{p\pm}/\zeta_0+\gamma_\pm \right)}
\nonumber \\
=-{e\Delta_0\over \pi\hbar}\frac{2 \sqrt{z_{p\pm}(2+z_{p\pm})}(1+z_{p\pm})}{4+4z_{p\pm}+{2iL\over \zeta_0}\sqrt{(1+4z_{p\pm}+2z_{p\pm}^2)^2-1}}
\nonumber \\
=-{e\Delta_0\over 2\pi \hbar}\frac{\zeta_0\sqrt{z_{p\pm}(2+z_{p\pm})}}{iL\sqrt{z_{p\pm}(2+z_{p\pm})}+\zeta_0}\approx {iev_f\over 2\pi  L}.
       \eem{res}
The contribution of poles to the current,
\be
J_{R\pm}=-2\pi \mbox{Im}\sum_p{\cal R}_{p\pm}=-{ev_f\over   L}p_\pm,
      \ee{resL}
is determined by the numbers $p_\pm$ of poles for rightmovers and leftmovers.
Corrections to the total current $J_{R+}-J_{R-}$  given by this expression  are not more than of  the order of $\propto 1/L^3$. This is shown in Appendix \ref{aA}.

 All poles must have  $x_{p\pm}$ coordinates satisfying inequalities  $0<x_{p\pm}<x_m$. This imposes the condition on pole numbers $p$:
\be
{\gamma_\pm\over 2\pi}<p< {L\over \pi \zeta_0}\left(x_m+{\zeta_0\gamma_\pm\over 2 L}\right).
   \ee{}
One can divide the right-hand side of this inequality onto  an integer and a fractional part:
\be
{L\over \pi \zeta_0}\left(x_m+{\zeta_0\gamma_\pm\over 2L}\right) =p_{m\pm}+\alpha_{m\pm},
   \ee{pm}
where the integer $p_{m\pm}$ is chosen so that $0< \alpha_{m\pm}<1$. If both $\gamma_\pm=2\pi\alpha\mp \theta_0 $ are positive the numbers of poles are $p_\pm =p_{m\pm}$. But when 
 $\gamma_+$ becomes negative, an additional pole with $p=0$ appears for rightmovers, and $p_+=p_{m+}+1$. 

The contribution of the contour segment on the imaginary axis is
\bem
J_{i\pm}= - {e\Delta_0\over \pi\hbar}\mbox{Re}\int_0^{y_m}  \frac{2\sqrt{2iy-y^2}(1+iy) idy}{4iy-2y^2+1-\cos(2iLy/\zeta_0+ \gamma_\pm)} 
\nonumber \\
\approx -{e\Delta_0\zeta_0^{3/2}\over \pi\hbar L^{3/2}}
\mbox{Re}\int_0^\infty \frac{\sqrt{iy'}}{1-\cosh y'\cos \gamma_\pm+i\sin y'\sin \gamma_\pm} idy'
 \nonumber \\
 =- {e\Delta_0\zeta_0^{3/2}\over \sqrt{2}\pi\hbar L^{3/2}}
\int_0^\infty \frac{\sinh y'\sin \gamma_\pm-1+\cosh y'\cos \gamma_\pm}{(\cosh y'-\cos \gamma_\pm)^2} \sqrt{y'}dy'.
  \eem{}
Here a new integration variable $y'= 2L y/\zeta_0$ was introduced, and the limits  $y_m\to \infty$ and $L \to \infty$ were considered.   Corrections to this value of $J_{i\pm}$ are on the order of $1/L^{5/2}$.

The  current $J_{i+}$ diverges in the limit $\gamma_+\to 0$. In this limit 
\be
J_{i+}\approx  - {\sqrt{2}e\Delta_0\zeta_0^{3/2}\over \pi\hbar L^{3/2}}
\int_0^\infty \frac{y'^2-\gamma_+^2+2\gamma_+y'}{(y'^2+\gamma_+^2)^2} \sqrt{y'}dy'
=- {e\Delta_0\zeta_0^{3/2}\over \hbar L^{3/2}\sqrt{\gamma_+}}\mbox{H}(\gamma_+).
  \ee{ipm}

 There is also a contribution from the contour segment along the vertical path parallel to the imaginary axis with $x=x_m$:
\bem
J_{m\pm}=-{e\Delta_0\over \pi\hbar}\mbox{Re}\int_0^{y_m} \frac{4x_m^2\,idy}{4x_m^2-e^{2Ly/\zeta_0-i(2Lx_m/\zeta_0 +\gamma_\pm)} }
 \nonumber \\
=-{e\Delta_0\over \pi\hbar}\mbox{Re}\int_0^{y_m} \frac{4x_m^2\,idy}{4x_m^2-e^{2Ly/\zeta_0-2\pi i\alpha_{m\pm}} }. 
     \eem{}
The maximum of the integrand is at the crossing  point of  the contour path and the curve, on which poles lie (dashed line in Fig.~\ref{CP}). Its coordinates are
\be 
x_0=x_m,~~y_0\approx {\zeta_0\over L}\ln(2x_m).
     \ee{} 
After introducing the new integration variables $y'=y-y_0$, the integral becomes 
\bem
J_{m\pm}={e\Delta_0\over \pi\hbar}\mbox{Re}\int_{-y_0}^{y_m-y_0} \frac{idy'}{1-e^{{2Ly'\over\zeta_0}-2\pi i\alpha_{m\pm}} }
\nonumber \\
\approx {e\Delta_0\over \pi\hbar}\mbox{Re}\int_{-\infty}^\infty \frac{idy'}{1-e^{{2Ly'\over\zeta_0}-2\pi i\alpha_{m\pm}} }
   \nonumber \\
=-{ev_f\over 2\pi L}
 \left. \arctan \frac{\sin (2\pi \alpha_{m\pm})}{e^{2Ly'\over\zeta_0}-\cos (2\pi \alpha_{m\pm})}\right\rvert_{y'=-\infty}^{\tilde y}
  \nonumber \\
-{ev_f\over 2\pi L}
 \left. \arctan \frac{\sin (2\pi \alpha_{m\pm})}{e^{2Ly'\over\zeta_0}-\cos (2\pi \alpha_{m\pm})}\right\rvert_{y'=\tilde y}^\infty
=-{ev_f\over  L}\left(\alpha_{m\pm}-{1\over 2}\right),
     \eem{}
  where $\tilde y=(\zeta_0 /2L)\ln\cos (2\pi \alpha_{m\pm}) $ is the value of $y$ in the zero of the denominator in the arctan argument. We divided the integration interval on two  parts with $y<\tilde y$
  and $y>\tilde y$ in order to stay at the integration at the same branch of the multivalued $\arctan$ function. 
  
  Collecting all contributions together one obtains
 \bem
 J_\pm=J_{R\pm}+J_{m\pm}+J_{i\pm}=-{ev_f\over  L}\left(p_\pm+\alpha_{m\pm}-{1\over 2}\right)+J_{i\pm}
 \nonumber \\
   =-{ev_f\over  L}[p_{m\pm}+\alpha_{m\pm}+\mbox{H}(-\gamma_\pm)]+J_{i\pm}
\nonumber \\
   =-{e\Delta_0\over  \pi \hbar}x_m-{ev_f\over  L}\left({\gamma_\pm\over 2\pi }-{1\over 2}\right)-{ev_f\over  L}\mbox{H}(-\gamma_\pm)+J_{i\pm}.
      \eem{cont}
  
  The total continuum vacuum current is
 \bem
J_{vC} =J_+-J_- ={ev_f\over L}\left[ {\theta_0\over \pi}- \mbox{H}(\theta_0-2\pi \alpha )\right] +J_{i+}-J_{i-}.
      \eem{vc}
Because of importance and nontriviality of  the continuum vacuum current, the main term $\propto 1/L$ in this current is calculated in the Appendix \ref{aB} by another method.

The chain of poles close to the real axis corresponds to the chain of peaks of the transmission probability (transmission resonances). There is the parity effect for transmission resonances  similar to that for bound  Andreev states.   At tuning of the phase $\theta_0$  the resonances can move in and out of the continuum changing the number of resonances from odd to even and {\em vice versa}. Any crossing of the energy $\Delta_0$ by a resonance produces a current jump.

Resonances can also cross the energy corresponding to the cutoff $x_m$ at small variation of this arbitrarily chosen parameter. This would also change the parity of the number of resonances and produce a current jump. But any such crossing leads to a jump of the incommensurability parameter $\alpha_m$ from 0 to 1 or from 1 to 0. This produce a jump in the value of the integral $J_{m\pm}$ over the contour segment parallel to the imaginary axis [\eq{cont}] completely compensating the jump from crossing the high energy cutoff by a resonance. Eventually, only crossings of the energy $\Delta_0$ by a resonance produce current jumps.

The abrupt jumps of the continuum vacuum current of the order $1/L$  at $\theta_0=2\pi\alpha$ are equal in magnitude but opposite in  sign to similar jumps of the vacuum current in bound states. The same is true for the corrections  $\propto 1/L^{3/2}$. This follows from analytical expressions Eqs.~(\ref{jAs}) and (\ref{ipm}) for currents in bound and continuum states in the limit of small $\gamma_+$. For not small  $\gamma_+$, the analytical expression for the  term    $J_{i+}-J_{i-}  \propto 1/L^{3/2}$ in \eq{vc} was not found and it was calculated numerically with Mathematica.   Its value is equal in magnitude and opposite in sign to the $\propto 1/L^{3/2}$ term in the current in bound states with high accuracy. The relative difference between two absolute values is about $10^{-8}$.

\subsection{Total vacuum current}

\begin{figure}[!t] 
\centering
\includegraphics[width=0.40 \textwidth]{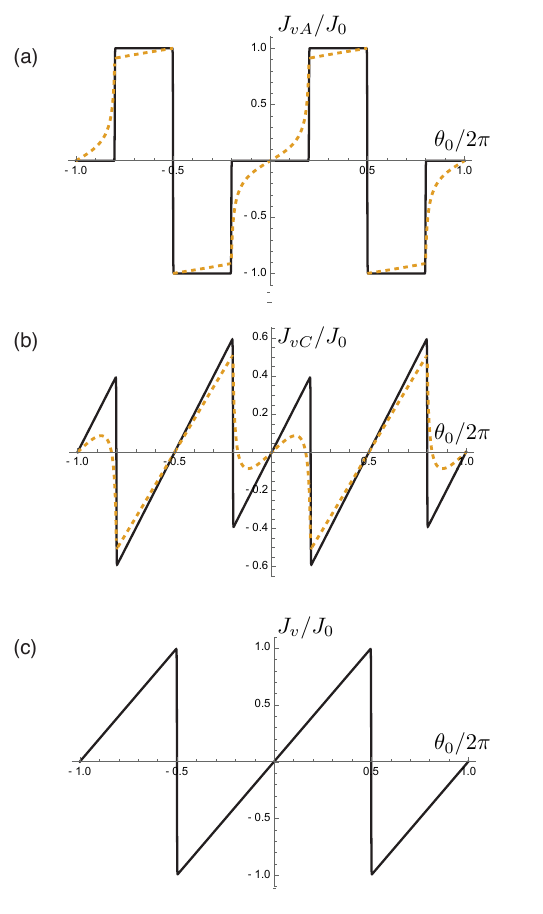} 
 \caption{The vacuum current in Andreev and continuum states at $\alpha=0.2$ in the 1D case. Solid lines show currents calculated taking into account only the main terms $\propto 1/L$. Currents calculated taking into account also corrections $\propto 1/L^{3/2}$ at $L/\zeta_0=25$  are shown by dashed lines. Here $J_0=ev_f/L$.  (a) The vacuum current $J_{vA}$ in Andreev states.  (b) The vacuum current $J_{vC}$ in continuum states.  (c) The  total vacuum current $J_v=J_{vA}+J_{vC}$. \label{vacC}}
 \end{figure}

The vacuum current $J_{vA}$ in Andreev states  and the continuum vacuum current $J_{vC}$ in the 1D case are shown in Figs.~\ref{vacC}(a) and \ref{vacC}(b) respectively.  Figure~\ref{vacC}(c) shows the total vacuum current  $J_v=J_{vA}+J_{vC}$.  Currents calculated neglecting or taking into account corrections $\propto 1/L^{3/2}$ are shown by solid or dashed lines, respectively.
The abrupt jumps of currents  produced by the parity effect  in the limit $L\to \infty$ are broadened by corrections on the order of $1/L^{3/2}$, which transform them into steep slopes with kinks exactly in the point $\theta_0=2\pi\alpha$ \cite{Son21}. These jumps and corrections $\propto 1/L^{3/2}$ in the bound states and the continuum  compensate each other in the total current, and  solid and dashed lines merge. Finally, the plot $J_v$ vs. $\theta_0$ is the same  saw-tooth relation as obtained at $T=0$ (Fig.~\ref{ST})  when the condensate current is the only current in the junction:
\bem
J_v={ ev_f\over L}{\theta_0\over \pi}= {en\hbar \over 2m }{\theta_0\over  L},
    \eem{v1d}
where $n=2k_f/\pi$ is the 1D electron density.

This picture agrees with calculations at finite $L$ by Bagwell \cite{Bagw} [see his Fig.~5(a)].\footnote{Currents in this figure have a sign opposite to that in our Fig.~\ref{vacC}. Apparently,  Bagwell took into account the negative electron charge. In the present paper the electron charge is included into $J_0$, and the plots  show ratios of currents to $J_0$ independent from the electron charge.} Bagwell revealed kinks of currents  in  the bound states and in the continuum and their mutual compensation in the total current. But he did not connect  them with  the parity effect and the incommensurability parameter $\alpha$, which were not considered by him.

Respective  contributions of the bound states and  the continuum  to the total vacuum current depend  on dimensionality \cite{Son21}. 
The transition from 1D to multidimensional systems requires averaging of currents  over the incommensurability parameter $\alpha$. After averaging the current $J_{vC}$ in continuum states 
and corrections $\propto 1/L^{3/2}$ both in the bound states and the continuum vanish, and the current $J_{vA}$ in bound states becomes equal to the total current $J_v$ described by the saw-tooth relation shown in Fig.~\ref{vacC}(c). 

 For multidimensional systems currents calculated for a single 1D channel  must be integrated over the space of wave vectors transverse to the current direction keeping in mind that
\be
v_f={\hbar k_f\over m}={\hbar \sqrt{k_F^2-k_\perp^2}\over m},
       \ee{}
where $k_F$ is the Fermi radius of a multidimensional system and $k_\perp$ is a transverse wave vector. The integration operation is $\int_{-k_F}^{k_F}  {dk_\perp\over 2\pi}… $ in the 2D  case and $\int_0^{k_F}  {k_\perp\,dk_\perp\over 2\pi}… $ in the 3D case. After integration one obtains that the expression  \eq{v1d} for the current derived for the 1D case is valid also for 
multidimensional systems if the 1D electron density $n$ is replaced by the 2D or 3D electron densities.

\subsection{Excitation current}

For  temperatures much higher than the Andreev level energy spacing (small $\beta$), one can replace the sum in  \eq{JqIn} by an integral. Its value yields the excitation current
\bem
J_q =-{ev_f\over L} {\theta\over \pi}. 
    \eem{JqEx}

\begin{figure}[h]%
\centering
\includegraphics[width=0.5\textwidth]{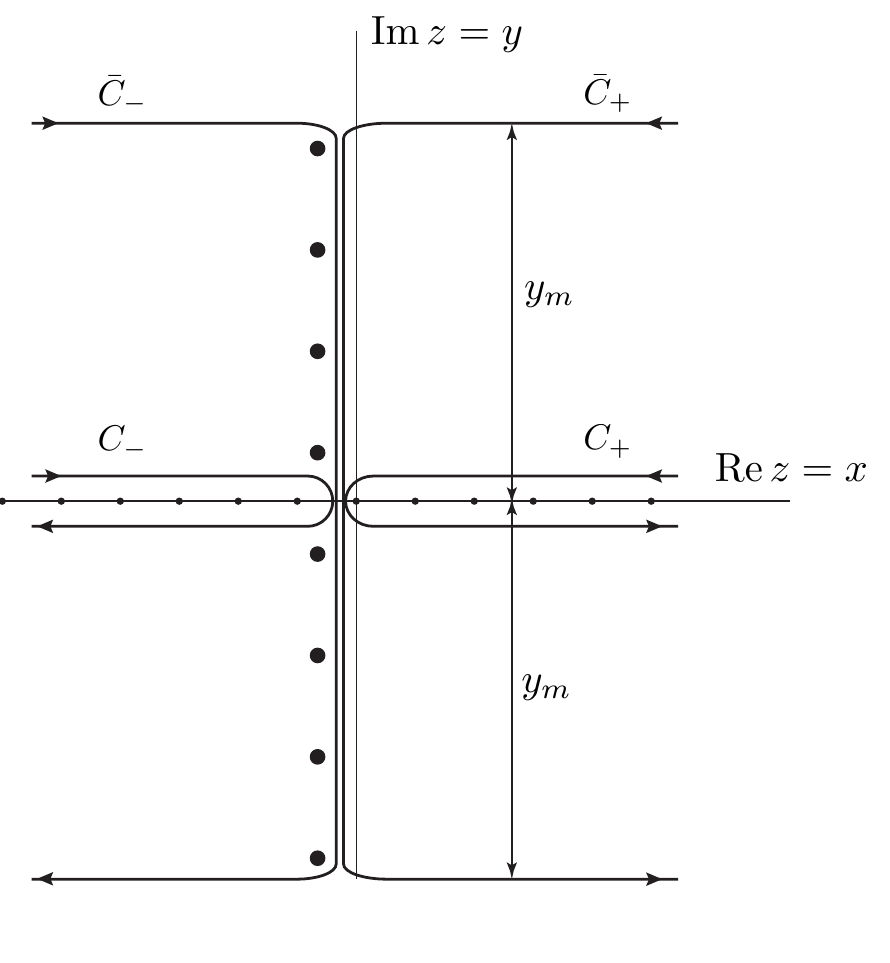}
\caption{Contour integration in the complex plane $z=x+iy$ for the calculation of the excitation current given by \eq{JqIn}. Small black circles show integer values of $x=s$, which numerate terms of the original sums in  \eq{JqSt}.  Large black circles show Matsubara poles.} \label{f2}
\end{figure}

In order to find  corrections  to this approximate expression we transform the  sum in \eq{JqIn}  to
\bem
J_q= {2ev_f\over L}  \left[\sum_{s=0}^\infty\frac{1}{e^{\beta \left(s+{\pi+\theta \over 2\pi}\right)}+1}-\sum_{s=-\infty}^{-1}\frac{1}{e^{-\beta \left(s+{\pi+\theta \over 2\pi}\right)}+1}\right].
      \eem{JqSt}
Next one can use the Matsubara approach transforming the sum to integrals over  two contours $C_+$ and $C_-$ in the complex plane $s=z=x+iy$ shown in Fig.~\ref{f2}: 
\be
J_q={ev_f\over  i L} \int_{C_+} \frac{\cot (\pi z)}{e^{\beta \left(z+{\pi+\theta \over 2\pi}\right)}+1}\,dz
+{ev_f\over iL} \int_{C_-} \frac{\cot(\pi z)}{e^{-\beta \left(z+{\pi+\theta \over 2\pi}\right)}+1}\,dz.
    \ee{JqC}
Then we perform analytical continuation transforming the contours $C_+$ and $C_-$ to the contours $\bar C_+$ and $\bar C_-$ (Fig.~\ref{f2}). At this transformation  the Matsubara poles
on the vertical line $x=-{1\over 2}$, 
\be
z_t=-{1\over 2} -{\theta \over 2\pi}+{2\pi i\over \beta}\left(t+{1\over 2}\right),
 \ee{}
are crossed, and the current is 
\be
J_q={ev_f\over iL}\left[ \int\limits_{\bar C_+} \frac{\cot(\pi z)}{e^{\beta \left(z+{\pi+\theta \over 2\pi}\right)}+1}\,dz
+\int\limits_{\bar C_-}\frac{\cot (\pi z)}{e^{-\beta \left(z+{\pi+\theta \over 2\pi}\right)}+1}\,dz\right]+\sum_{t=-\infty}^\infty M_t, 
    \ee{JqCb}
where $M_t$ is the contribution of the $t$th pole. 

Main contributions from contour integrals come from 4 horizontal segments of paths $\bar C_+$ and $\bar C_-$. They are integrals over real $x$ at large constant imaginary parts $iy=\pm iy_m$. The contribution   of the upper segment of contour $\bar C_+$ is
\bem
{ev_f\over  iL}\int _\infty^{-{1\over 2}}\frac{\cot[\pi( i y_m +x)]}{e^{\beta \left(i y_m +x+{\pi+\theta \over 2\pi}\right)}+1}\,dx
\approx-{ev_f\over  L}\int _\infty^{-{1\over 2}}\frac{dx}{e^{\beta \left(i y_m +x+{\pi+\theta \over 2\pi}\right)}+1}
\nonumber \\
 =\left. {ev_f\over  L\beta} \ln\left[ 1+e^{-\beta\left(i y_m +x+{\pi+\theta \over 2\pi}\right)}\right]\right\rvert_{x=\infty}^{-{1\over 2}}
={ev_f\over  L\beta} \ln\left[ 1+e^{-\beta \left(iy_m+ {\theta \over 2\pi}\right)}\right].
    \eem{}

Making similar estimations of other three horizontal segments one obtains the total contribution of all four horizontal segments, 
\be
{ev_f\over  L\beta}\ln\left\lvert\frac{1+e^{\beta \left(iy_m- {\theta \over 2\pi}\right)}}{1+e^{\beta \left(iy_m+{\theta \over 2\pi}\right)}}\right\rvert^2
={ev_f\over  L\beta}\ln\frac{1+2\cos (\beta y_m)e^{- {\beta\theta \over 2\pi}}+e^{- {\beta\theta \over \pi}}}{1+2\cos (\beta y_m)e^{{\beta\theta \over 2\pi}}+e^{{\beta\theta \over \pi}}},
     \ee{}
which  in the limit $\beta \to 0$ yields the excitation current $J_q$ given by \eq{JqEx}. The total contribution of vertical segments of the contours $\bar C_+$ and $\bar C_-$ vanishes. 

The contribution of the $t$th Matsubara pole is
\be
M_t={2  \pi ev_f\over \beta L}\frac{\tan {\theta\over 2} -i  \tanh\left[{2\pi^2 \over \beta}\left(t+{1\over 2}\right)\right]}
 {1+i \tan {\theta\over 2} \tanh\left[{2\pi ^2\over \beta}\left(t+{1\over 2}\right)\right]} .
    \ee{}
In the high temperature limit $\beta\to 0$ only the poles with $t=0$ and $t=-1$ are important. Finally, the excitation current is
\bem
J_q=-{ev_f\over \pi L}  \theta+M_0+M_{-1}=-{ev_f\over \pi L}  \theta+{8e T\over \hbar } e^{-2\pi TL/\hbar v_f}\sin \theta.
     \eem{JqM}
The sinusoidal term in this equation was known before \cite{Bard}. 

Integration over transverse wave vectors in multidimensional systems yields 
\bem
J_q= -en{\hbar \over 2m }{\theta\over  L}+A_q en \sqrt{T\over m k_F L}e^{-2\pi TL/\hbar v_F}\sin \theta,
   \eem{}
where
\be
A_q=\left\{\begin{array}{cc}
8  ~~ & \mbox{2D case} \\ &
 \\
6\pi~~  & \mbox{3D case} \end{array} \right. .
     \ee{Aq}

\section{Current-phase relation}

The current-phase relation   is determined by the condition that the sum of the vacuum and  the excitation current vanishes.  At zero temperature the both current vanish, and the current-phase relation is the saw-tooth curve shown in Fig.~\ref{ST}. In the 1D case at high temperatures the condition imposed by the charge conservation law is
\bem 
J_v+J_q=en{\hbar \over 2m }{\theta_0\over  L}-en{\hbar \over 2m }{\theta \over  L}+{8e T\over \hbar } e^{-2\pi TL/\hbar v_f}\sin \theta
\nonumber \\
=-en{\hbar \over 2m }{\theta_s \over  L}+{8e T\over \hbar } e^{-2\pi TL/\hbar v_f}\sin \theta=0.
     \eem{Jvq}
The current-phase relation is
\be
J=J_s=en{\hbar \over 2m }{\theta_s \over  L}={8e T\over \hbar } e^{-2\pi TL/\hbar v_f}\sin \theta
    \ee{1D}
for the 1D system, and 
\bem 
J=A_q en \sqrt{T\over m k_F L}e^{-2\pi TL/\hbar v_F}\sin \theta
     \eem{mD}
for multidimensional systems. The constant $A_q$ is given by \eq{Aq}.

Because the total current $J$ is exponentially decreases with growing $L$ and $T$, the SNS junction becomes a weak link,  and the widely accepted in the past approach  neglecting phase gradients in superconducting leads gives the correct result. The phase $\theta_s$ connected with phase gradients is small, and  the difference between the Josephson phase $\theta=\theta_0+\theta_s $ and the vacuum phase $\theta_0$ is not important. Ignoring $\theta_s$ in  \eq{Jvq}, the current $J_v+J_q$ yields a correct value of the total current $J$ despite formal violation of the charge conservation law. 

The exponentially small current at high temperature is a result of mutual cancellation of the large terms $\propto 1/L$ in the vacuum current $J_v$ and the temperature independent part of the excitation current $J_q$. It is not evident that this cancellation is exact. Any power law correction $\propto 1/L^w$  independent from temperature becomes more important than the exponentially decaying current at temperatures exceeding the temperature
\be
 T^*={\hbar v_F\over 2\pi L }\ln{L  \over C},  
     \ee{tem}
where the value of $C$ weakly (logarithmically) depends on $T$ and $L$. Due to a large logarithmic factor in \eq{tem},  the temperature $T^*$ is much larger than the Andreev level energy spacing but, nevertheless, is much smaller than the gap $\Delta_0$, and all assumptions made in the calculations are valid.  

\section{Discussion and conclusions} \label{DiC}

The present paper investigates the long ballistic SNS junction using the approach initiated in Ref.~\cite{Son21}. The approach focused on the problem with the charge conservation law  in  the self-consistent field method dealing with the effective pairing potential that breaks the gauge invariance. This requires to filter solutions obtained by this method keeping only those, which do not violate the charge conservation law. The filtration is provided by the condition  that the total currents inside all three layers must be equal. The total current consists of three parts: $J=J_s+J_v +J_q$. The condensate current $J_s=env_s$  is produced by the phase gradient in the superconducting layers and is the same in all three layers and, therefore, it does not violate the charge conservation law. The vacuum current $J_v$ and the excitation current $J_q$ flow only in the normal layer, and the charge conservation law requires that $J_v+J_q=0$. The current-phase relation of the junction is derived from  this condition. 

At zero temperature the new approach \cite{Son21}  yielded the same saw-tooth current-phase relation as previous investigations (Fig.~\ref{ST}). But its physical picture was different. 
  In the previous investigations the current $J_s$ was ignored, and sloped segments were related only with the vacuum current  $J_v$ (although the adjective ``vacuum'' was not used). This was in conflict with the charge conservation law, which requires that the current $J_v$ does not flow alone.  According to the new approach, slope segments of the curve correspond to the motion of the electron fluid as a whole, and the condensate current $J_s$ is the only current in the junction, the value of which  is simply derived from the Galilean invariance for Andreev scattering.   
  The calculation of $J_v$ using  the sophisticated formalism of finite temperature Green's functions \cite{Kulik,Ishii} is not relevant for the current-phase relation at $T=0$.
  At $T=0$ the electron transport in the SNS junctions does not differ from that in a uniform superconductor. 

At nonzero temperatures, one should determine the current-phase relation from the condition $J_v+J_q=0$, and calculations of two currents $J_v$ and $J_q$ are necessary. At the calculation of the vacuum current $J_v$ in  Ref.~\cite{Son21} the vacuum current in continuum states was neglected. It was correct for  multidimensional (2D and 3D) systems, but not for the 1D case. The present paper reports a  calculation of the vacuum current in continuum states correcting this error and retracting the prediction of Ref.~\cite{Son21} that the SNS junction can be a $\varphi_0$ junction.

The new calculation of the vacuum current in continuum states showed that the parity effect revealed for the 1D case in  Ref.~\cite{Son21} for the vacuum current in bound states exists also in continuum states. The parity effect produces current jumps and kinks on the current-phase plots. These jumps in bound states and in the continuum are equal in magnitude and opposite in sign, and in the total vacuum current two parity effects compensate each other. In the past current jumps and kinks were revealed in calculations of the current-phase relation in the 1D case by Bagwell \cite{Bagw}, but their connection with the parity effect was not realized.

One might ask why an essential difference in the physical pictures does not lead to a difference in the  outcome of the analysis. This ``insensitivity'' to the physical picture existed already in the past. The physical pictures of  Ishii \cite{Ishii} and  Bardeen and Johnson \cite{Bard} are not the same. Bardeen and Johnson \cite{Bard} assumed that the excitation current $J_q$ and the condensate current $J_s$ derived from the Galilean invariance flow  in the normal layer (ignoring that the same current $J_s$  flows also in the superconducting layers). Ishii \cite{Ishii} assumed that   the excitation and the vacuum currents $J_q$ and $J_v$ flow in the normal layer at $T=0$. However, the vacuum current $J_v= {ev_f\over L}{\theta_0\over \pi}$  and the condensate current   $J_s= {ev_f\over L}{\theta_s\over \pi}$ have the same expressions as function of their phases, while the  excitation current at high temperatures $J_q= -{ev_f\over L}{\theta \over \pi}$ (neglecting the exponentially small term from Matsubara poles) has the similar expression via its phase but with an opposite sign.  Since Ishii \cite{Ishii}  and  Bardeen and Johnson \cite{Bard} ignored differences between phases $\theta_0$ and $\theta$ it did not matter  whether $J_s$ or   $J_v$ flows in the normal layer. 
  
Coincidence of the expressions for different currents as functions of their phases takes place in the approximation taking into account only the main terms $\propto 1/L$ and $\propto 1/L^{3/2}$ in the $1/L$ expansion. The agreement with the results of previous works using \cite{Ishii} or not using \cite{Bard}  the formalism of Green's functions points out that their results were obtained in the same approximation. The {\em ab initio} expressions for different currents are not identical, and the cancellation of temperature independent terms  in $J_v+J_q$ at high temperatures might  not retain in a more accurate  approximation taking into account  terms decreasing with $L$ faster than  $ 1/L^{3/2}$.  The estimation of these terms is a subject for a future investigation.  
 
Recently Thuneberg \cite{ThunKink} presented a numerical calculation of the currents in the bound states and the continuum. He also revealed kinks on the current-phase dependences, which compensate each other in the total current. This agrees with numerical calculation of these currents by Bagwell \cite{Bagw} and with our analytical calculation connecting the kinks with the parity effect (odd vs. even number of Andreev levels).

\bmhead{Acknowledgments}

Interactions and discussions with Alexander Andreev were very important for my research activity in superfluidity and magnetism. He supported my suggestion of spin superfluidity \cite{ES-78b} despite it was rejected by some influential Moscow theoreticians. This support was not obtained as granted, and I had to spend a number of nights in the train St. Petersburg (then Leningrad) - Moscow for meetings with Andreev to discuss this issue.

\begin{appendices}
 
\section{Corrections to the contribution of poles in the continuum vacuum current}\label{aA}

Let us expand the expression \eq{res} for residues of poles in $1/L$:
\bem
{\cal R}_{p\pm}\approx {iev_f\over 2\pi  L}\left[1+ {i\zeta_0\over L\sqrt{z_{p\pm}(2+z_{p\pm})} }
-{\zeta_0^2\over L^2z_{p\pm}(2+z_{p\pm})}\right].
       \eem{}
We need only imaginary parts of residues which give real contributions to the current. Using the expressions for coordinates of poles in the complex plane $z=x+iy$ given by \eq{poles} and the inequality $y_{p\pm}\ll  x_{p\pm}$ we obtain the  correction to the main term  $\propto 1/L$:
\bem
\mbox{Im}\sum_p\delta {\cal R}_{p\pm}={ev_f\zeta_0\over 2\pi  L^2}\sum_p\left[{y_{p\pm}(x_{p\pm}+1)\over (2x_{p\pm}+x_{p\pm}^2)^{3/2}}- {\zeta_0\over L (2x_{p\pm}+x_{p\pm}^2)}\right]
\nonumber \\
={ev_f\zeta_0^2\over 2\pi  L^3}\sum_p\left[{(x_{p\pm}+1)\ln(\sqrt{2x_{p\pm}+x_{p\pm}^2}+1+x _{\pm p} )\over (2x_{p\pm}+x_{p\pm}^2)^{3/2}}-{1\over 2x_{p\pm}+x_{p\pm}^2}\right]. 
   \eem{}
Terms in the sum smoothly depend on $x_{p\pm}$ without singularities, and one can replace summation by integration over continuous $p$. Since $dp =L dx_{p\pm}/\pi\zeta_0$ the integral is on the order of $1/L^2$. But the integral does not depend on $\theta_0$ and vanishes in the difference $\mbox{Im}\sum_p\delta {\cal R}_{+ p}-\mbox{Im}\sum_p\delta {\cal R}_{- p}$, which determines the total current of  rightmovers and leftmovers. In order to separate the part depending on $\theta_0$ one must first to take a derivative in $\theta_0$ of any term in the sum first and to replace summation by integration afterwards. Since $d\theta_0=2L dx_{p\pm}/\zeta_0$ the $\theta_0$ dependent correction is not more than of the order $1/L^3$.

\section{Calculation of the continuum vacuum current
by averaging over transmission oscillations}\label{aB}

Here we present another method of calculation of the main term $\propto 1/L$ in the continuum vacuum current, which helps to better understand the role of transmission resonances. 
We  divide the whole interval of integration in the integral \eq{jvc} on intervals of the length equal to the oscillation period  $\pi\zeta_0/L$ of the integrand:
\bem
J_\pm=-{e\Delta_0\over \pi \hbar}\left[I_0(\gamma_\pm)+ \sum_{p=1}^{p_{m\pm}} I(p,\gamma_\pm)+I_m(\gamma_\pm) \right],
     \eem{jcp}
where
\be
I(p,\gamma_\pm)=\int_{-\pi\zeta_0/2L}^{\pi\zeta_0/2L}  \frac{2\sqrt{2z_{p\pm}+z_{p\pm}^2}(1+z_{p\pm})}{4z_{p\pm}+2z_{p\pm}^2+1-\cos(2Lz'/\zeta_0)}\,dz'={\pi\zeta_0\over L}.
   \ee{is1}
Here $z_{p\pm}=x_{p\pm}$ is the real coordinate of the transmission resonance given  by \eq{poles},      $z'=z-z_{p\pm}$, and $p_{m\pm}$ is the maximal integer $p$ satisfying the condition that the upper border of the $p_m$th  period does not exceed some large $x_m$, which in the end must go to infinity: 
 \be
 z_{p_m\pm}+{\pi\zeta_0\over 2L} = {(2\pi p_{m\pm}+\pi-\gamma_\pm)\zeta_0\over 2L}<x_m. 
     \ee{}
   Inside of any interval we neglected variation of the variable $z$ excepting its variation   in the argument of the cosine function. Everywhere else we replace $z$ by its value $z_{p\pm}=x_{p\pm}$ in the interval center, which coincides with location of the $p$th transmission resonance. The sum of terms $I(p,\gamma_\pm)$ is
\be
 \sum_{p=1}^{p_{m\pm}} I(p,\gamma_\pm)={\pi\zeta_0\over L} p_{m\pm}.
      \ee{}

The term  $I_m(\gamma_\pm)$ takes into account the integration interval between the upper limit $x_m$ in the integral \eq{jvc} and  the upper border of the $p_{m\pm}$th  period. The integrand at large $z$ goes to 1, and
\be
I_m(\gamma_\pm)= x_m- z_{p_m\pm}-{\pi\zeta_0\over 2L}= x_m-  {(2\pi p_{m\pm}+\pi-\gamma_\pm)\zeta_0\over 2L}. 
     \ee{} 
 
 The term  $I_0(\gamma_\pm)$ takes into account the integration interval between $z=0$ and  the lower  border of the $p=1$ period. In the limit $L\to \infty$ the integrand at small $z$ can be approximated by a delta function:
\bem
 I_0(\gamma_\pm)=\int_0^{(\pi-\gamma_\pm)\zeta_0/2L}\frac{2\sqrt{2z}\,dz}{4z+1-\cos(2Lz/\zeta_0+ \gamma_\pm)}
 \nonumber \\
 \approx \int_0^{(\pi-\gamma_\pm)\zeta_0/2L}{\pi \zeta_0\over L}\delta\left(z-{\gamma_\pm\zeta_0\over 2L}\right) dz={\pi \zeta_0\over L}\times\left\{  \begin{array}{cc}  0
& \mbox{at}~\gamma_\pm>0 \\
\\
1 & \mbox{at}~\gamma_\pm<0 \end{array}    \right..
    \eem{} 
 
 Substituting the expression for $I_0(\gamma_\pm)$, $ \sum\limits_{p=1}^{p_{m\pm}} I(p,\gamma_\pm)$, and $I_m(\gamma_\pm)$ into \eq{jcp} one obtains the value of $J_\pm$ identical to the term $\propto 1/L$ in \eq{cont}.
 
\end{appendices}

%\bibliography{bibBook,nano,spinB}



\begin{thebibliography}{27}
\ifx \bisbn   \undefined \def \bisbn  #1{ISBN #1}\fi
\ifx \binits  \undefined \def \binits#1{#1}\fi
\ifx \bauthor  \undefined \def \bauthor#1{#1}\fi
\ifx \batitle  \undefined \def \batitle#1{#1}\fi
\ifx \bjtitle  \undefined \def \bjtitle#1{#1}\fi
\ifx \bvolume  \undefined \def \bvolume#1{\textbf{#1}}\fi
\ifx \byear  \undefined \def \byear#1{#1}\fi
\ifx \bissue  \undefined \def \bissue#1{#1}\fi
\ifx \bfpage  \undefined \def \bfpage#1{#1}\fi
\ifx \blpage  \undefined \def \blpage #1{#1}\fi
\ifx \burl  \undefined \def \burl#1{\textsf{#1}}\fi
\ifx \doiurl  \undefined \def \doiurl#1{\url{https://doi.org/#1}}\fi
\ifx \betal  \undefined \def \betal{\textit{et al.}}\fi
\ifx \binstitute  \undefined \def \binstitute#1{#1}\fi
\ifx \binstitutionaled  \undefined \def \binstitutionaled#1{#1}\fi
\ifx \bctitle  \undefined \def \bctitle#1{#1}\fi
\ifx \beditor  \undefined \def \beditor#1{#1}\fi
\ifx \bpublisher  \undefined \def \bpublisher#1{#1}\fi
\ifx \bbtitle  \undefined \def \bbtitle#1{#1}\fi
\ifx \bedition  \undefined \def \bedition#1{#1}\fi
\ifx \bseriesno  \undefined \def \bseriesno#1{#1}\fi
\ifx \blocation  \undefined \def \blocation#1{#1}\fi
\ifx \bsertitle  \undefined \def \bsertitle#1{#1}\fi
\ifx \bsnm \undefined \def \bsnm#1{#1}\fi
\ifx \bsuffix \undefined \def \bsuffix#1{#1}\fi
\ifx \bparticle \undefined \def \bparticle#1{#1}\fi
\ifx \barticle \undefined \def \barticle#1{#1}\fi
\bibcommenthead
\ifx \bconfdate \undefined \def \bconfdate #1{#1}\fi
\ifx \botherref \undefined \def \botherref #1{#1}\fi
\ifx \url \undefined \def \url#1{\textsf{#1}}\fi
\ifx \bchapter \undefined \def \bchapter#1{#1}\fi
\ifx \bbook \undefined \def \bbook#1{#1}\fi
\ifx \bcomment \undefined \def \bcomment#1{#1}\fi
\ifx \oauthor \undefined \def \oauthor#1{#1}\fi
\ifx \citeauthoryear \undefined \def \citeauthoryear#1{#1}\fi
\ifx \endbibitem  \undefined \def \endbibitem {}\fi
\ifx \bconflocation  \undefined \def \bconflocation#1{#1}\fi
\ifx \arxivurl  \undefined \def \arxivurl#1{\textsf{#1}}\fi
\csname PreBibitemsHook\endcsname

%%% 1
\bibitem{And64}
\begin{barticle}
\bauthor{\bsnm{Andreev}, \binits{A.F.}}:
\batitle{The thermal conductivity of the intermediate state in
  superconductors}.
\bjtitle{Zh. Eksp. Teor. Fiz.}
\bvolume{46},
\bfpage{1823}--\blpage{1828}
(\byear{1964}).
\bcomment{[{\em Sov. Phys.--JETP}, {\bf 19}, 1228--1231 (1964)]}
\end{barticle}
\endbibitem

%%% 2
\bibitem{Lif57}
\begin{barticle}
\bauthor{\bsnm{Lifshitz}, \binits{E.M.}},
\bauthor{\bsnm{Pitaevskii}, \binits{L.P.}}:
\batitle{Absorption of second sound in rotating helium {II}}.
\bjtitle{Zh. Eksp. Teor. Fiz.}
\bvolume{33},
\bfpage{535}--\blpage{537}
(\byear{1957}).
\bcomment{[{\em Sov. Phys.--JETP}, {\bf 6}, 418--419 (1958)]}
\end{barticle}
\endbibitem

%%% 3
\bibitem{Son75}
\begin{barticle}
\bauthor{\bsnm{Sonin}, \binits{E.B.}}:
\batitle{Friction between the normal component and vortices in rotating
  superfluid helium}.
\bjtitle{Zh. Eksp. Teor. Fiz.}
\bvolume{69},
\bfpage{921}--\blpage{935}
(\byear{1975}).
\bcomment{[{\em Sov. Phys.--JETP}, {\bf 42}, 469--475 (1976)]}
\end{barticle}
\endbibitem

%%% 4
\bibitem{Gal}
\begin{barticle}
\bauthor{\bsnm{Gal'perin}, \binits{Y.M.}},
\bauthor{\bsnm{Sonin}, \binits{E.B.}}:
\batitle{Motion of vortices and electrical conductivity of pure type {II}
  supercomductors in weak magnetic fields}.
\bjtitle{Fiz. Tverd. Tela (Leningrad)}
\bvolume{18},
\bfpage{3034}--\blpage{3041}
(\byear{1976}).
\bcomment{[{\em Sov. Phys.--Solid State}, {\bf 18}, 1768--1772 (1976)]}
\end{barticle}
\endbibitem

%%% 5
\bibitem{Kop76}
\begin{barticle}
\bauthor{\bsnm{Kopnin}, \binits{N.B.}},
\bauthor{\bsnm{Kravtsov}, \binits{V.E.}}:
\batitle{Forces acting on vortices moving in a pure type {II} superconductor}.
\bjtitle{Zh. Eksp. Teor. Fiz.}
\bvolume{71},
\bfpage{1644}--\blpage{1656}
(\byear{1976}).
\bcomment{[{\em Sov. Phys.--JETP}, {\bf 44}, 861--867 (1976)].}
\end{barticle}
\endbibitem

%%% 6
\bibitem{EBS}
\begin{bbook}
\bauthor{\bsnm{Sonin}, \binits{E.B.}}:
\bbtitle{Dynamics of Quantised Vortices in Superfluids}.
\bpublisher{Cambridge University Press},
\blocation{Cambridge}
(\byear{2016})
\end{bbook}
\endbibitem

%%% 7
\bibitem{Lancast}
\begin{barticle}
\bauthor{\bsnm{Tsepelin}, \binits{V.}},
\bauthor{\bsnm{Baggaley}, \binits{A.W.}},
\bauthor{\bsnm{Sergeev}, \binits{Y.A.}},
\bauthor{\bsnm{Barenghi}, \binits{C.F.}},
\bauthor{\bsnm{Fisher}, \binits{S.N.}},
\bauthor{\bsnm{Pickett}, \binits{G.R.}},
\bauthor{\bsnm{Jackson}, \binits{M.J.}},
\bauthor{\bsnm{Suramlishvili}, \binits{N.}}:
\batitle{Visualization of quantum turbulence in superfluid $^3${He}-${B}$:
  Combined numerical and experimental study of {A}ndreev reflection}.
\bjtitle{Phys. Rev. B}
\bvolume{96},
\bfpage{054510}
(\byear{2017}).
\doiurl{10.1103/PhysRevB.96.054510}
\end{barticle}
\endbibitem

%%% 8
\bibitem{ScrbSerg}
\begin{barticle}
\bauthor{\bsnm{Skrbek}, \binits{L.}},
\bauthor{\bsnm{Sergeev}, \binits{Y.A.}}:
\batitle{Feasibility of an analog of {A}ndreev reflection in superfluid
  $^{4}\mathrm{He}$}.
\bjtitle{Phys. Rev. B}
\bvolume{108},
\bfpage{100502}
(\byear{2023}).
\doiurl{10.1103/PhysRevB.108.L100502}
\end{barticle}
\endbibitem

%%% 9
\bibitem{And65}
\begin{barticle}
\bauthor{\bsnm{Andreev}, \binits{A.F.}}:
\batitle{Electron spectrum of the intermediate state of superconductors}.
\bjtitle{Zh. Eksp. Teor. Fiz.}
\bvolume{49},
\bfpage{655}--\blpage{660}
(\byear{1964}).
\bcomment{[{\em Sov. Phys.--JETP}, {\bf 22}, 455--458 (1966)]}
\end{barticle}
\endbibitem

%%% 10
\bibitem{Kulik}
\begin{barticle}
\bauthor{\bsnm{Kulik}, \binits{I.O.}}:
\batitle{Macroscopic quantization and proximity effect in {S--N--S} junctions}.
\bjtitle{Zh. Eksp. Teor. Fiz.}
\bvolume{57},
\bfpage{1745}--\blpage{1759}
(\byear{1969}).
\bcomment{[Sov. Phys.--JETP, {\bf 30}, 944--950 (1970)]}
\end{barticle}
\endbibitem

%%% 11
\bibitem{Ishii}
\begin{barticle}
\bauthor{\bsnm{Ishii}, \binits{C.}}:
\batitle{Josephson currents through junctions with normal metal barriers}.
\bjtitle{Prog. Theor. Phys. (Japan)}
\bvolume{44},
\bfpage{1525}--\blpage{1546}
(\byear{1970})
\end{barticle}
\endbibitem

%%% 12
\bibitem{Bard}
\begin{barticle}
\bauthor{\bsnm{Bardeen}, \binits{J.}},
\bauthor{\bsnm{Johnson}, \binits{J.L.}}:
\batitle{Josephson current flow in pure
  {S}uperconducting-{N}ormal-{S}uperconducting junctions}.
\bjtitle{Phys. Rev B}
\bvolume{5},
\bfpage{72}--\blpage{78}
(\byear{1972})
\end{barticle}
\endbibitem

%%% 13
\bibitem{deGen}
\begin{bbook}
\bauthor{\bparticle{de} \bsnm{Gennes}, \binits{P.G.}}:
\bbtitle{Superconductivity of Metals and Alloys}.
\bpublisher{Benjamin},
\blocation{New York}
(\byear{1966})
\end{bbook}
\endbibitem

%%% 14
\bibitem{Thun}
\begin{barticle}
\bauthor{\bsnm{Thuneberg}, \binits{E.}}:
\batitle{Comment on ``{B}allistic {SNS} sandwich as a {J}osephson junction''}.
\bjtitle{Phys. Rev. B}
\bvolume{108},
\bfpage{176501}
(\byear{2023})
\end{barticle}
\endbibitem

%%% 15
\bibitem{Son21}
\begin{barticle}
\bauthor{\bsnm{Sonin}, \binits{E.B.}}:
\batitle{Ballistic {SNS} sandwich as a {J}osephson junction}.
\bjtitle{Phys. Rev. B}
\bvolume{104},
\bfpage{094517}
(\byear{2021})
\end{barticle}
\endbibitem

%%% 16
\bibitem{Bagwell}
\begin{barticle}
\bauthor{\bsnm{Riedel}, \binits{R.A.}},
\bauthor{\bsnm{Chang}, \binits{L.-F.}},
\bauthor{\bsnm{Bagwell}, \binits{P.F.}}:
\batitle{Critical current and self-consistent order parameter of a
  superconductor--normal-metal--superconductor junction}.
\bjtitle{Phys. Rev. B}
\bvolume{54},
\bfpage{16082}--\blpage{16095}
(\byear{1996}).
\doiurl{10.1103/PhysRevB.54.16082}
\end{barticle}
\endbibitem

%%% 17
\bibitem{Sols}
\begin{barticle}
\bauthor{\bsnm{Sols}, \binits{F.}},
\bauthor{\bsnm{Ferrer}, \binits{J.}}:
\batitle{Crossover from the {J}osephson effect to bulk superconducting flow}.
\bjtitle{Phys. Rev. B}
\bvolume{49},
\bfpage{15913}--\blpage{15919}
(\byear{1994}).
\doiurl{10.1103/PhysRevB.49.15913}
\end{barticle}
\endbibitem

%%% 18
\bibitem{diode}
\begin{botherref}
\oauthor{\bsnm{Davydova}, \binits{M.}},
\oauthor{\bsnm{Prembabu}, \binits{S.}},
\oauthor{\bsnm{Fu}, \binits{L.}}:
Universal {J}osephson diode effect.
Sci. Adv.
\textbf{8}(23)
(2022).
\doiurl{10.1126/sciadv.abo0309}
\end{botherref}
\endbibitem

%%% 19
\bibitem{Son23rep}
\begin{barticle}
\bauthor{\bsnm{Sonin}, \binits{E.B.}}:
\batitle{Reply to {C}omment on ``{B}allistic {SNS} sandwich as a {J}osephson
  junction''}.
\bjtitle{Phys. Rev. B}
\bvolume{108},
\bfpage{176502}
(\byear{2023})
\end{barticle}
\endbibitem

%%% 20
\bibitem{Buzdin}
\begin{barticle}
\bauthor{\bsnm{Buzdin}, \binits{A.}}:
\batitle{Direct coupling between magnetism and superconducting current in the
  {J}osephson ${\ensuremath{\varphi}}_{0}$ junction}.
\bjtitle{Phys. Rev. Lett.}
\bvolume{101},
\bfpage{107005}
(\byear{2008}).
\doiurl{10.1103/PhysRevLett.101.107005}
\end{barticle}
\endbibitem

%%% 21
\bibitem{Svidz71}
\begin{barticle}
\bauthor{\bsnm{Svidzinskii}, \binits{A.V.}},
\bauthor{\bsnm{Antsygina}, \binits{T.N.}},
\bauthor{\bsnm{Bratus}, \binits{E.N.}}:
\batitle{Superconducting current in wide {S--N--S} junctions}.
\bjtitle{Zh. Eksp. Teor. Fiz.}
\bvolume{61},
\bfpage{1612}--\blpage{1619}
(\byear{1971}).
\bcomment{[Sov. Phys.--JETP, {\bf 34}, 860--863 (1972)]}
\end{barticle}
\endbibitem

%%% 22
\bibitem{Kummel}
\begin{barticle}
\bauthor{\bsnm{Gunsenheimer}, \binits{U.}},
\bauthor{\bsnm{Sch\"ussler}, \binits{U.}},
\bauthor{\bsnm{K\"ummel}, \binits{R.}}:
\batitle{Symmetry breaking, off-diagonal scattering, and {J}osephson currents
  in mesoscopic weak links}.
\bjtitle{Phys. Rev. B}
\bvolume{49},
\bfpage{6111}--\blpage{6125}
(\byear{1994}).
\doiurl{10.1103/PhysRevB.49.6111}
\end{barticle}
\endbibitem

%%% 23
\bibitem{Tin}
\begin{bbook}
\bauthor{\bsnm{Tinkham}, \binits{M.}}:
\bbtitle{Introduction to Superconductivity},
\bedition{2}nd edn.
\bpublisher{McGrow-Hill},
\blocation{New York}
(\byear{1996})
\end{bbook}
\endbibitem

%%% 24
\bibitem{Bagw}
\begin{barticle}
\bauthor{\bsnm{Bagwell}, \binits{P.F.}}:
\batitle{Suppression of the {J}osephson current through a narrow, mesoscopic,
  semiconductor channel by a single impurity}.
\bjtitle{Phys. Rev. B}
\bvolume{46},
\bfpage{12573}--\blpage{12586}
(\byear{1992}).
\doiurl{10.1103/PhysRevB.46.12573}
\end{barticle}
\endbibitem

%%% 25
\bibitem{5}
\begin{bbook}
\bauthor{\bsnm{Gradshteyn}, \binits{I.S.}},
\bauthor{\bsnm{Ryzhik}, \binits{I.M.}}:
\bbtitle{Table of Integrals, Series, and Products},
\bedition{7}th edn.
\bpublisher{Academic Press},
\blocation{Amsterdam}
(\byear{2007})
\end{bbook}
\endbibitem

%%% 26
\bibitem{ThunKink}
\begin{botherref}
\oauthor{\bsnm{Thuneberg}, \binits{E.}}:
Square-well model for superconducting pair-potential
(2024).
\url{https://doi.org/10.48550/arXiv.2405.07659}
\end{botherref}
\endbibitem

%%% 27
\bibitem{ES-78b}
\begin{barticle}
\bauthor{\bsnm{Sonin}, \binits{E.B.}}:
\batitle{Analogs of superfluid currents for spins and electron-hole pairs}.
\bjtitle{Zh. Eksp. Teor. Fiz.}
\bvolume{74},
\bfpage{2097}--\blpage{2111}
(\byear{1978}).
\bcomment{[Sov. Phys.--JETP, {\bf 47}, 1091--1099 (1978)]}
\end{barticle}
\endbibitem

\end{thebibliography}


\end{document}